\documentclass[onecolumn]{mn2e}  
\usepackage{mnras_cite}  
 \usepackage[dvips]{graphicx}   
\usepackage{amsmath}
\newcommand{\nc}{\newcommand}    

\nc{\de}{\delta} 
\nc{\tISW}{\triangle_T^{ISW}}
\nc{\hn}{\hat{n}}
\nc{\bH}{\bar{H}} 
\nc{\Ol}{\Om_{\Lambda}} 

\nc{\ul}{\underline} \nc{\al}{\alpha} \nc{\g}{\gamma}
\nc{\Del}{\Delta} \nc{\e}{\textrm{e}} \nc{\eps}{\epsilon}
\nc{\lam}{\lambda} \nc{\Om}{\Omega} \nc{\Omm}{\Omega_m}
\nc{\Oml}{\Omega_\Lambda} \nc{\LCDM}{$\Lambda$CDM~} 
\nc{\ve}{\varepsilon} \nc{\mn}{{\mu\nu}} \nc{\vp}{\varphi}

\def\gsim{\; \raise0.3ex\hbox{$>$\kern-0.75em
\raise-1.1ex\hbox{$\sim$}}\; }

\nc{\Section}[2]{\section{#2}\label{#1}}    
\nc{\Bibitem}[1]{\bibitem{#1}}    
\nc{\Label}[1]{\label{#1}}    
 
\nc{\beq}[1]{\begin{equation}\label{#1}}      
\nc{\eeq}{\end{equation}}

\nc{\hq}{\hat{q}}
\nc{\hw}{\widehat{w}}

\def\ben{\begin{enumerate}}
\def\een{\end{enumerate}}
\def\bi{\begin{itemize}}
\def\ei{\end{itemize}}
\def\ee{\end{equation}}
\def\bea{\begin{eqnarray}}
\def\eea{\end{eqnarray}}

\nc{\Mpc}{Mpc/h}    
\nc{\vev}[1]{\langle #1 \rangle}    
    
\def\ltsima{$\; \buildrel < \over \sim \;$}    
\def\gtsima{$\; \buildrel > \over \sim \;$}    
\def\simlt{\lower.5ex\hbox{\ltsima}}    
\def\simgt{\lower.5ex\hbox{\gtsima}}    
\nc{\w}{$w_2(\theta)$\ }    
\nc{\ie}{i.e.}     
\nc{\eg}{e.g.}

\input epsf

\usepackage{amsmath}
\usepackage{amssymb}

\def\xisp{\xi(\sigma, \pi)}
\def\xis{\xi(s)}
\def\xir{\xi(r)}

\nc{\be}[1]{\begin{equation}\mbox{$\label{#1}$}
             
            \end{equation}
}

\begin{document}

\title{Clustering of luminous red galaxies II:\\
small scale redshift space distortions}

\author[Cabr\'e \& Gazta\~{n}aga]{Anna Cabr\'e, Enrique Gazta\~{n}aga\\ 
Institut de Ci\`encies de l'Espai, CSIC/IEEC, Campus UAB, F. de Ci\`encies, Torre C5 par-2, Barcelona 08193, Spain}
  
  \twocolumn

\maketitle


\begin{abstract} 
This is the second paper of a series where we study 
the clustering of LRG galaxies in the latest spectroscopic SDSS 
data release, DR6, which has 75000 LRG galaxies covering over 1 $Gpc^3/h^3$
for $0.15<z<0.47$.   
Here we focus on modeling 
redshift space distortions in $\xisp$,
the 2-point correlation in separate line-of-sight and 
perpendicular directions, at small scales and in the line-of-sight.
We show that a simple Kaiser
model for the anisotropic 2-point correlation function in redshift space,
convolved with a distribution of random peculiar velocities with an exponential
form, can describe well the correlation of LRG at all scales. We show that to
describe with accuracy the so called "fingers-of-God" (FOG) elongations in the
radial direction, it is necessary to model the
scale dependence of both bias $b$ and the pairwise rms peculiar velocity
$\sigma_{12}$ with the distance. We show how both quantities can be inferred 
from the $\xisp$ data. From $r \simeq 10$ Mpc/h to $r \simeq 1$ Mpc/h, both the bias 
and $\sigma_{12}$ are shown to increase by a factor of two: from $b=2$ to $b=4$ and from
$\sigma_{12}=400$ to $800$ Km/s. The later is in good
agreement, within a 5 percent accuracy
in the recovered velocities, with direct velocity measurements
in dark matter simulations with $\Omega_m=0.25$ and $\sigma_8$=0.85.\
\end{abstract}    

\maketitle    

    
\section{Introduction}    
\label{sec:intro}

The luminous red galaxies (LRGs) are selected by color and magnitude to obtain
intrinsically red galaxies in Sloan Digital Sky Survey (SDSS)
\cite{eisenstein2001}. These galaxies trace a big volume, around
$1Gpc^3h^{-3}$, which make them perfect to study large scale clustering
or smaller scales with larger statistics.
In this paper we focus on the 2-point correlation function $\xisp$ on 
the smaller scales, where the non-linear bias and
the random peculiar velocities can complicate the analysis. Our motivation is 
to provide a model that can explain the observed $\xisp$ of the LRG galaxies.
Such a model does not exist right now and we are not aware of
any attempt to reproduce the $\xisp$ in the detail we will explore here.
LRG galaxies seem to display larger "Fingers-of-God" \cite{jackson} than regular
galaxies, is this evidence for larger velocities? If so, is this
evidence that LRG trace stronger gravitational potentials?
We will show that a simple Kaiser
model for the anisotropic 2-point correlation function in redshift space,
convolved with a distribution of random peculiar velocities with an exponential
form, can describe well the correlation of LRG at all the scales. To
describe with accuracy the so called "fingers-of-God" (FOG) elongations in the
radial direction, it is necessary to model the
scale dependence of both bias $b$ and the pairwise rms peculiar velocity
$\sigma_{12}$ with the distance.

In Paper I \cite{paper1} of this series, we have analyzed 
larger scales and present the basis for this work, including 
more detailed theory, error and systematic effects.

On large scales, the density fluctuations are small enough to be linearized, and
can be used to constrain cosmological parameters, since we can assume that the
clustering is well described by (linearly biased) dark matter 
(see Paper I). On smaller
scales, we can learn about the relation of galaxies to dark matter through the
biased clustering of halos. This range of distances can be fitted by a power
law, but there are small deviations that can be understood in the theory of the
halo occupation distribution. The transition between galaxy pairs of the same
halo and galaxy pairs that belong to different halo, occurs around 1Mpc/h. When
moving to scales smaller than 1Mpc/h, in the 1-halo term, we can see processes
more complex that modify the galaxy clustering, such as dynamical friction,
tidal interactions, stellar feedback, and other dissipative processes.

We use the most recent spectroscopic SDSS data release, DR6
\cite{dr6}, to perform the study of small scales in the anisotropic 2-point
correlation function, and its derivatives projected correlation function and
real-space correlation function. LRGs are supposed to
be red old elliptical galaxies, which are usually passive galaxies, with
relatively low star formation rate. They have steeper slopes in the correlation
function than the rest of galaxies, since they are supposed to reside in the
centers of big halos, inducing non-linear bias dependent on scale, for small
scales.

The same LRGs (but with reduced area) have been studied from different points of
view. 
Zehavi et al (2005)
study LRGs at intermediate scales (0.3 to 40Mpc/h),
where they calculate the projected correlation function, the monopole and
real-space correlation function to study mainly linear bias,
non-linear bias and the differences between luminosities. They find that there
are differences from a power law for scales smaller than 1Mpc/h and they find no strong evolution for LRG.
At smaller
scales, 
Eisenstein et al (2005)
did a cross-correlation between spectroscopic LRG with
photometric main sample in order to reduce shot-noise in small scales
clustering. They conclude that the clustering is higher for most luminous galaxies. Moreover, LRG have a scale dependent bias on luminosity, while for normal galaxies the luminosity bias is scale independent. LRG galaxies are surrounded by other red galaxies near them, which seem to be approaching the center LRG galaxy. Finally, 
Masjedi et al (2006) deal with very small scale clustering to scales
smaller than 55'' by cross-correlating the spectroscopic LRG sample and the
targeted imaging sample and find that the correlation function from 0.01-8Mpc/h
is really close to a power law with slope -2, but there are still some features
that diverge from the power law.


 The small scale slope depends on the interplay
between two factors which control how the correlation function of galaxies is
related to that of the underlying matter : the number of galaxies within a dark
matter halo (HOD) and the range of halo masses which contain more than one
galaxy \cite{benson}. LRGs are to be found in halos with the median of the
distribution occurring at $3\;10^{13}M_\odot/h$, estimated using weak lensing
measurements \cite{mandelbaum}. 
Almeida et al (2008) find with simulations that 25\% of LRGs at z = 0.24 are satellite
galaxies, which play an important role in the form of the small scale correlation function, as well as
the pairwise
velocity dispersion, and also provide information about galaxy formation and
evolution.
Zheng et al (2008), who study HOD in LRG clustering, find that the satellite fraction is small (5-5\% for $M_g<-21.2$) and decreases with LRG luminosity.

Slosar et al (2006)
show that pairwise velocity distribution in real space is a
complicated mixture of host-satellite, satellite-satellite and two-halo pairs.
The peak value is reached at around 1Mpc/h and does not reflect the velocity
dispersion $\sigma_{12}$ of  a typical halo hosting these galaxies, but is instead dominated
by the sat-sat pairs in high-mass clusters. 
Different groups have found this dependence of $\sigma_{12}$ on the scale (Zehavi et al 2002, Hawkins et al 2003, Jing and Borner 2004, Li et al 2006, Van den Bosch 2007, Li et al 2007). They also find that $\sigma_{12}$ depends on luminosity, 
with higher values for both large and small luminosities,
but this tendency can not be predicted by the halo model. Li et al (2006, 2007) add that redder galaxies have stronger clustering and larger velocities at all scales, so they move in strongest gravitational fields.
Tinker et al (2007)
use the halo
occupation distribution framework to make robust predictions of the pairwise
velocity dispersion (PVD). They assume that central galaxies move with the
center of mass of the host halo and satellite galaxies move as dark matter. The
pairs that involve central galaxies have a lower dispersion, so the fraction of
satellites strongly influences both the luminosity and scale dependence of the
PVD in their predictions. At r between 1 and 2 Mpc/h, the PVD rapidly increases with smaller separation as
satellite-satellite one-halo pairs from massive halos dominate. At r $<$ 1 Mpc/h, the PVD
decreases towards smaller scales because central-satellite one-halo pairs become more
common. 
At r $>$ 3Mpc/h, the PVD is dominated by two-halo central galaxy pairs, and reach a constant value. 

LRGs have also been analyzed at higher redshifts (z=0.55) with the 2dF-SDSS LRG
and QSO Survey (2SLAQ, 
Cannon et al 2007). 
Ross et al (2007), Da \^Angela et al (2008) and Wake et al (2008)
analyze the
redshift distortions in the LRGs and quasars for this catalog. For part of our
analysis we have followed the method explained in 
Hawkins et al (2003), an extensive
analysis of the redshift distortions in the 2dF catalog.

Here we define the parameters that we assume during all this work, which are
motivated by recent results of WMAP, SNIa and previous LSS analysis: $n_s=0.98$,
$\Omega_b=0.045$, $h=0.72$ and flat geometry. Unless otherwise said, we use $\Omega_m=0.25$ and $\sigma_8=0.85$, both obtained from paper I. We will use the power spectrum analytical form for
dark matter by 
Eisenstein and Hu (1998), and the non-linear fit to halo theory by
Smith et al (2003).

\section{Modeling redshift-space}\label{sec:model}

We follow the modeling given in more detailed in Paper I \cite{paper1}.
Here we just show the main equations related to this work.
In the large-scale linear regime
and in the plane-parallel approximation (where galaxies are taken to be
sufficiently far away from the observer that the displacements induced by
peculiar velocities are effectively parallel), the distortion caused by coherent
infall velocities takes a particularly simple form in Fourier space \cite{kaiser}:

\begin{equation}\label{eq:kaiserpoint}
P_s(k) = (1 + \beta\mu_k^2)^2 P(k).
\end{equation}
where $P(k)$ is the power spectrum of density fluctuations $\delta$,
$\mu$ is the cosine of the angle between $k$ and the line-of-sight, 
the subscript $s$ indicates redshift space, and $\beta$ is the growth rate of growing modes in linear theory.

The observed value of $\beta$ is
\begin{equation} 
\label{fb}
 \beta = {f(\Omega_m) \over b} \equiv {1 \over b} {d\;ln\;D \over d\;ln\;a} 
\end{equation}
where b is the bias between galaxies and dark matter, 
$f(\Omega_m)$ and $D(\Omega_m)$ are the linear growth density
and velocity factors. Here we use
Hamilton (1992) who translated Kaiser results into real space
$\xi'(\sigma, \pi)$ (see Eq.(8) in Paper I).
We then convolve it with the distribution function of random
pairwise velocities, $f(v)$, to give the final model $\xisp$ \cite{peebles1980}:

\begin{equation}
\label{eq:hamiltonmethod}
\xisp = \int^{\infty}_{-\infty}\xi'(\sigma, \pi - v/H(z)/a(z))f(v)dv
\end{equation} 

We represent the random motions by an exponential form (Szalay et al, 1998),
\begin{equation}
f(v) = \frac{1}{\sigma_{12}\sqrt{2}}\exp\left(-\frac{\sqrt{2}|v|}{\sigma_{12}}\right)
\label{e:fv}
\end{equation} 

where $\sigma_{12}$ is the pairwise peculiar velocity dispersion. \\

It is well known that $\sigma_{12}$ depends on the real separation between particles $r=\sqrt{\sigma^2 + \pi_{real}^2}$, where $\pi_{real}=\pi - v/H(z)/a(z)$. The pairwise velocity dispersion is roughly constant for $r>5Mpc/h$, but increases towards smaller values of $r$. To use Eq.(\ref{eq:hamiltonmethod}) and Eq.(\ref{e:fv}) with $\sigma_{12}$ depending on $r$, we do the following. For each value of the velocity $v$ in the integral, we estimate the real
distance $r=\sqrt{\sigma^2 + \pi_{real}^2}$ and the value of $\sigma_{12}=\sigma_{12}(r)$ that enters in $f(v)$. 
We have checked that for $\sigma>2Mpc/h$, this exact method is very similar to assuming that $\sigma_{12}$ is fixed for each $\sigma$ and it is constant along the line-of-sight $\pi$, so we can change $\sigma_{12}$ depending on $\sigma$, rather than $r$. This has some practical advantages over the exact method.

We  define the multipoles of $\xisp$ as
\begin{equation}\label{eq:moment}
\xi_{\ell}(s) = \frac{2\ell+1}{2}\int^{+1}_{-1}\xisp P_{\ell}(\mu)d\mu.
\end{equation}
where $\mu$ is cosine of the angle to the line-of-sight $\pi$. The 
monopole $\xis \equiv \xi_0(s)$ is also called the redshift space
correlation function.
The real-space correlation function can be estimated from
the projected correlation function, $\Xi(\sigma)$, 
by integrating the redshift distorted $\xisp$ along the
 line-of-sight $\pi$:

\begin{equation}
\Xi(\sigma) = 2\int^{\infty}_{0}\xisp~d\pi
\label{e:Xi}
\end{equation}

Davis and Peebles (1983) show that $\Xi(\sigma)$ is directly related to the real-space correlation function.

\begin{equation}
\Xi(\sigma)=
2\int^{\infty}_{\sigma}\frac{r \xir dr}{(r^2 -
\sigma^2)^{\frac{1}{2}}}.
\label{e:proj}
\end{equation}

It is possible to estimate $\xir$ by directly inverting $\Xi(\sigma)$ \cite{saunders1992}. We can write Eq.(\ref{e:proj}) as,
\begin{equation}
\xir = -\frac{1}{\pi}\int^{\infty}_r \frac{(d\Xi(\sigma)/d\sigma)}{(\sigma^2 - r^2)^{\frac{1}{2}}}d\sigma.
\label{eq:xirr}
\end{equation}
Assuming a step function for $\Xi(\sigma)=\Xi_i$ in bins centered on
$\sigma_i$, and interpolating between values,
\begin{equation}
\label{eq:xirr2}
\xi(\sigma_i) = -\frac{1}{\pi}\sum_{j\geq i}\frac{\Xi_{j+1}-\Xi_j}{\sigma_{j+1}-\sigma_j}\ln\left(\frac{\sigma_{j+1}+\sqrt{\sigma^2_{j+1} - \sigma^2_i}}{\sigma_j + \sqrt{\sigma^2_j - \sigma^2_i}}\right)
\end{equation}
for $r = \sigma_i$. 

Once we recover the real-space correlation function, we can also estimate the
ratio of the redshift-space correlation function, $\xis$, to the real-space
correlation function, $\xir$, which gives an estimate of the redshift distortion
parameter, $\beta$, on large scales:

\begin{equation} \label{eq:xisr} \frac{\xis}{\xir} = 1 + \frac{2\beta}{3} +
\frac{\beta^2}{5}. 
\end{equation}

\section{Study of the errors}\label{sec:errors}

A detailed account of errors is given in Paper I.
Basically we use mock catalogs to estimate
what we call the Monte Carlo (MC) errors. 
Mock catalogs are build out of very large numerical 
simulations run  in the super computer Mare Nostrum in Barcelona
by MICE consortium (www.ice.cat/mice).
The  simulation contains $2048^3$ dark matter particles, in a cube of side
7680Mpc/h, $\Omega_M=0.25$, $\Omega_b =0.044$, $\sigma_8=0.8$, $n_s=0.95$ and
$h=0.7$. We use both dark matter and groups.
There is no bias in the dark matter mocks, so
$\beta=\frac{\Omega_m(z)^{0.55}}{b}=0.62$ (where $\Omega_m(z)=\Omega_m
(1+z)^3/(\Omega_m (1+z)^3+1-\Omega_m)$ and $b=1$). In this paper, we use group mocks with $M > 2.2\times10^{13} M_{\sun}$
(b = 1.9, $\beta = 0.25$) at z = 0, which are similar to real LRG galaxies in its clustering properties.

For the jackknife (JK)  error, we obtain the different realizations 
directly from the data, dividing the catalog in M zones, and we
consider that each realization is all the catalog except from one of these JK
zones. In this case, as the realizations are clearly not independent, we 
multiply the covariance matrix by a factor $(M-1)$ to account for this effect
(see Paper I for more details).

To analyze data we prefer to use a model independent error, such as the JK error
above. The MC error based on simulations depends on the model that we have used
and in particular in the overall normalization, which is similar to LRG for group simulations but can have slight differences.
We use the MC errors to probe the accuracy of JK errors in the same simulations.

In Paper I \cite{paper1}, we
present the errors in the redshift space correlation function $\xisp$ (bin=5Mpc/h),
the monopole $\xi(s)$ and the quadrupole $Q(s)$. Here we study the errors in the $\xisp$ (bin=1Mpc/h and 0.2Mpc/h),
perpendicular projected function $\Xi(\sigma)$ and in the real-space correlation
function $\xir$. We also use the simulations to test the validity of the
2-point correlation function model.

\subsection{Errors in the redshift space correlation $\xisp$}\label{sec:errorpisigma}

\begin{figure*}
	\centering{ \epsfysize=4.5cm\epsfbox{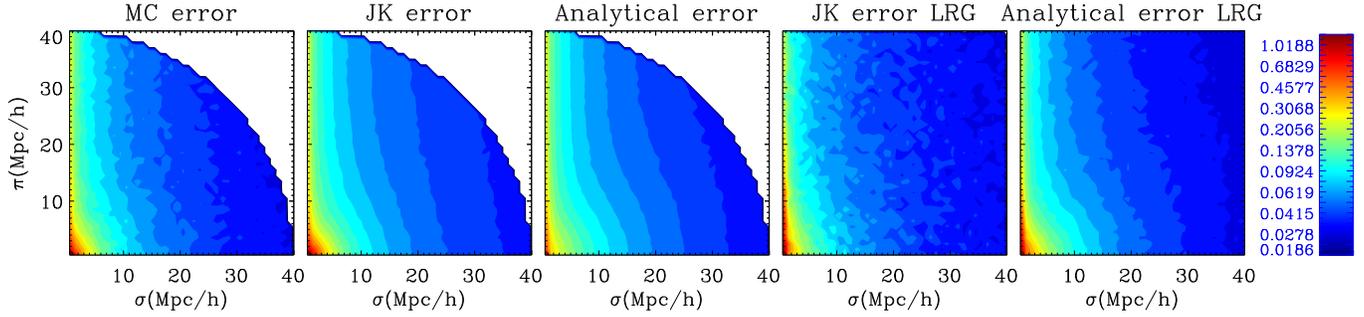}}
	\caption{We compare different error estimators for $\xisp$ with binning 1Mpc/h for MICE group mocks with b = 1.9: Monte Carlo, mean of all jackknife and analytical form. The last two plots refer to real LRG data: JK error and analytical form. Errors work well in the simulations, and they are similar in amplitude and shape to LRG real errors. 
	 \label{fig:err1Mpc}}
\end{figure*}

\begin{figure}
	\centering{ \epsfysize=6.5cm\epsfbox{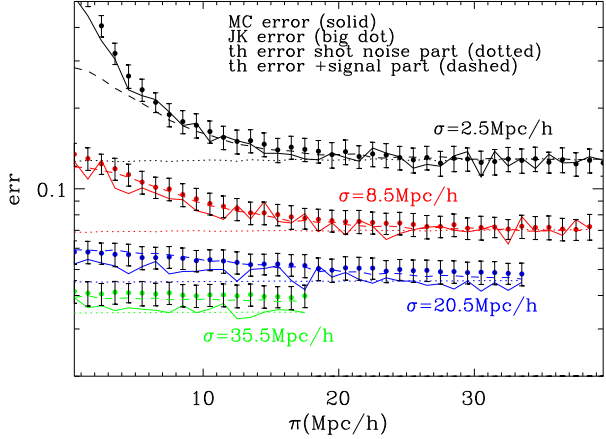}}
	\caption{We compare different error estimators for $\xisp$ with 1Mpc/h of bin for MICE group mocks with b = 1.9. We fix the perpendicular distance $\sigma$ and move along the line-of-sight $\pi$
as indicated in the figure. The errors are: MC (solid line), theory with (dashed) and without (dotted) the signal part of the error, and JK
(big dots with errorbars). As we move to higher $\sigma$ (lower amplitudes), JK starts to fail, but the analytical and MC errors there agree
well.
	 \label{fig:err21Mpc}}
\end{figure}

\begin{figure}
	\centering{ \epsfysize=4.5cm\epsfbox{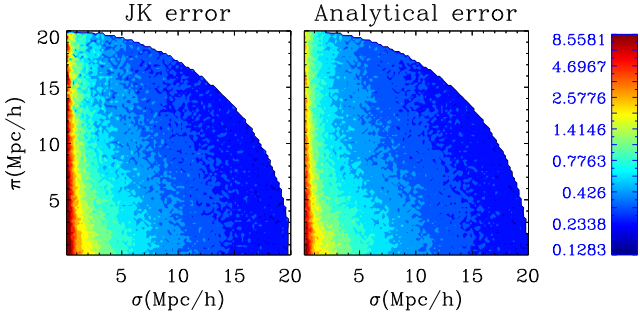}}
	\caption{We compare different error estimators for $\xisp$ with 0.2Mpc/h of binning for LRG real data: JK and analytical. As in the case of bin=1Mpc/h, JK seems to work for $\sigma<20Mpc/h$.
	 \label{fig:err02Mpc}}
\end{figure}

We calculate the 2-point correlation function in redshift space $\xisp$ for each group mock, using a bin of 1Mpc/h. We can obtain a JK error for each mock, so a mean and dispersion for all the mocks. We compare it to the MC error, and to the analytical form, which is described in detail in paper I. We propose the error to have the following form
$\Delta\xi= \Delta\xi_{shot-noise} + \Delta\xi_{signal} $, with
two arbitrary coefficients $\alpha_{noise}$ and $\alpha_{signal}$ , so that:

\begin{equation}
\Delta\xi= 
 \alpha_{noise} ~\Delta\xi_{Poisson} + \alpha_{signal} ~\xi
\label{eq:errth}
\end{equation}
 The comparison for the three kind of errors can be seen in Fig.\ref{fig:err1Mpc}, where it is also plotted the JK error for real LRG galaxies, and the corresponding analytical error. The amplitude and shape of the error is similar in the group mocks and in LRG data, so we assume that our conclusions derived from the simulations can be translated to real LRG data. In order to see the differences in the mock errors with more detail, we fix the perpendicular distance $\sigma$ for different cases and move along the line-of-sight $\pi$ in Fig.\ref{fig:err21Mpc}. The JK error starts to deviate slightly from MC for $\sigma$ higher than 20Mpc/h, where the shot noise analytical form works perfectly. For small $\sigma$ and $\pi$, the signal part of the analytical error helps to fit to MC error, but it seems that at these scales the best option is to use JK error. The covariance is lower than 0.2 for all the points, so it is nearly diagonal. In Fig.\ref{fig:err02Mpc} we compare the JK error and analytical error for real data for a bin of 0.2Mpc/h and we find similar conclusions. We have done the same analysis with dark matter simulations and they also work well, as expected.
Note that contrary to what we found in paper I for 5Mpc/h binning, for smaller bins the shot noise model matches the expectations with $\alpha_{noise}=1$, rather than $\alpha_{noise}=1.4$. We believe that this could be related to the smaller covariance in 0.2Mpc/h and 1Mpc/h binning.

\subsection{Errors in the projected correlation 
$\Xi(\sigma)$}\label{sec:errorprojected}

\begin{figure}
	\centering{ \epsfysize=6cm\epsfbox{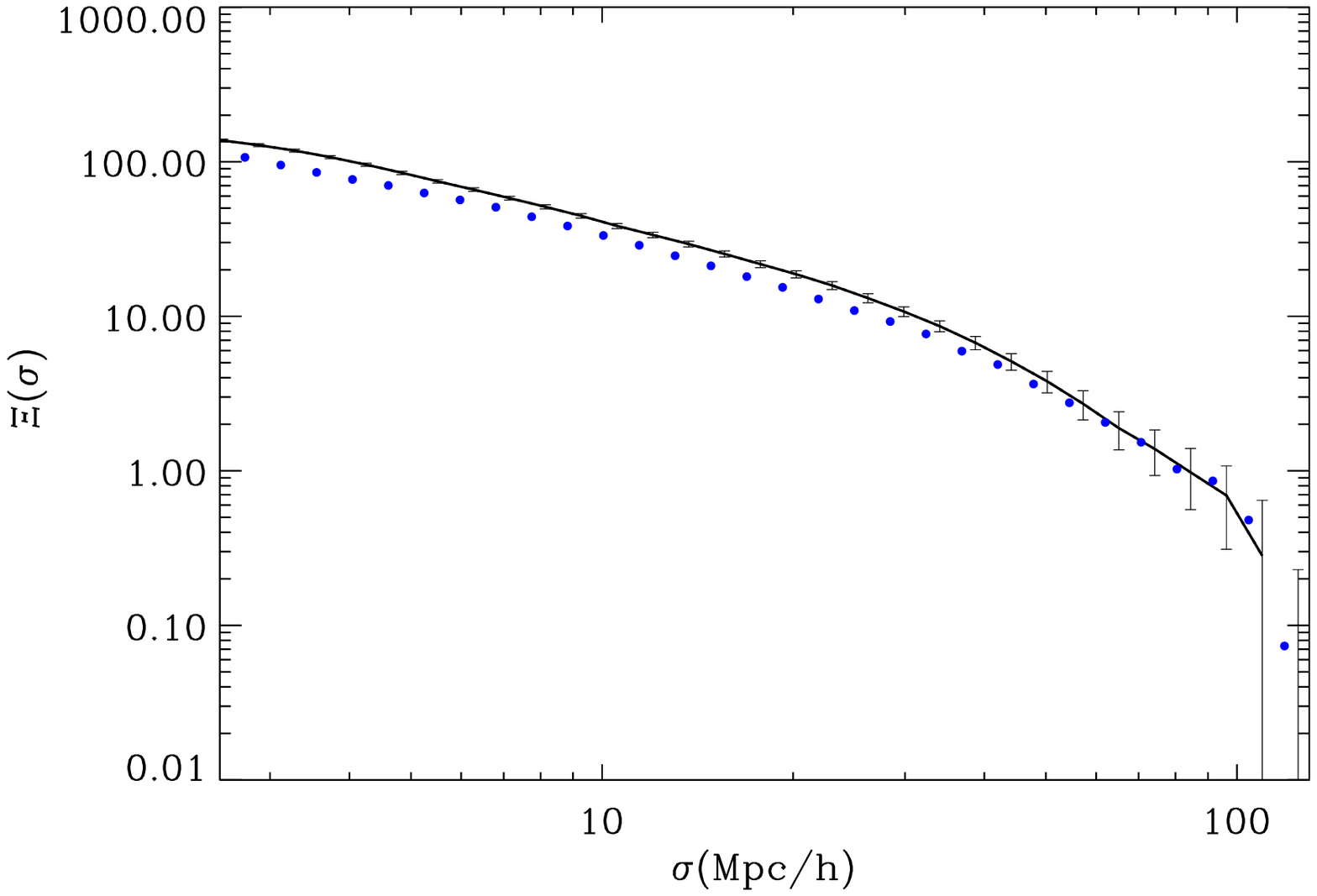}}
	\centering{ \epsfysize=6cm\epsfbox{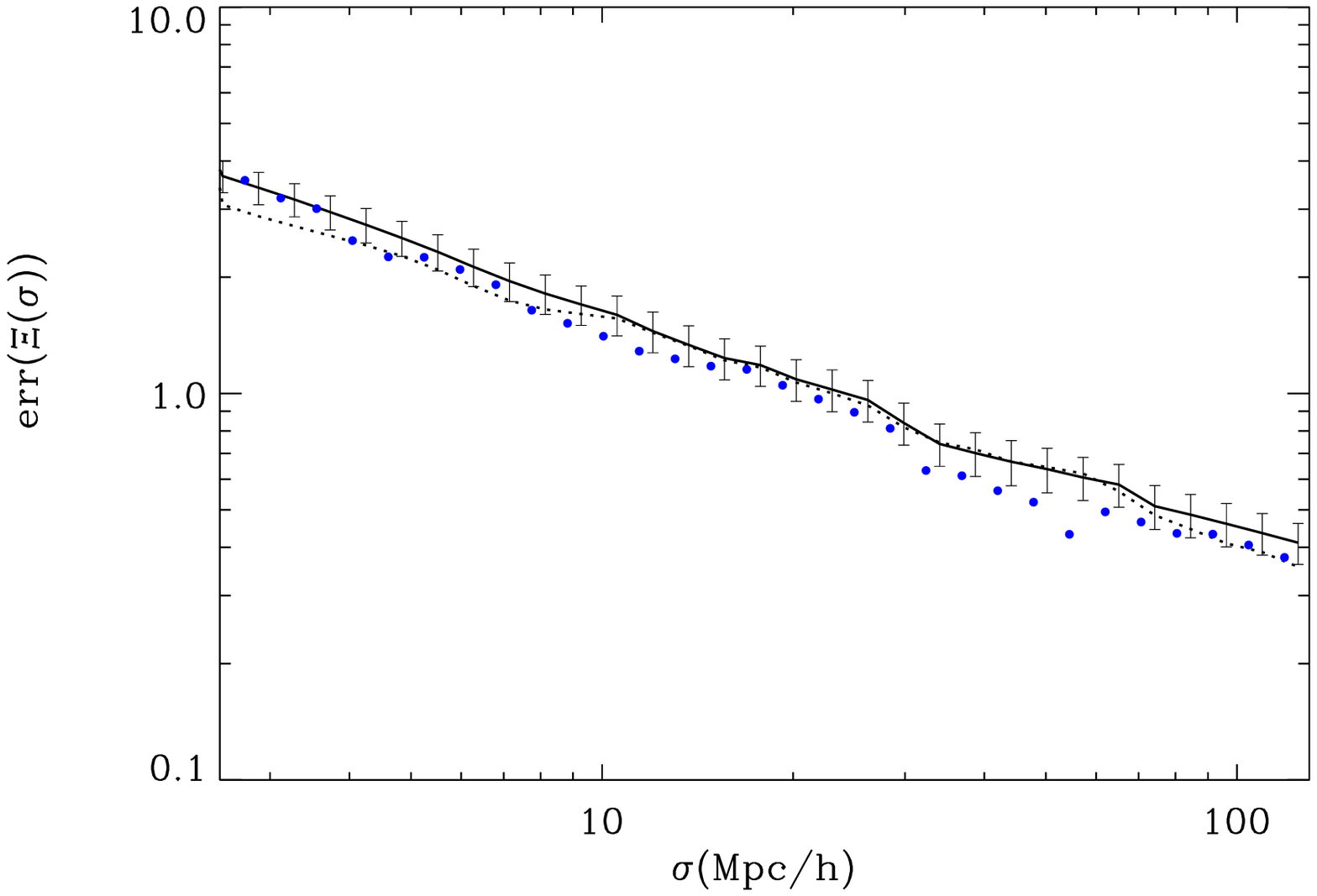}}
	\caption{Top: $\Xi(\sigma)$ for the group simulations calculated from 
	$\xisp$ (solid line with errors) and the value for LRG data (big dots).
	 Bottom: Errors in $\Xi(\sigma)$ for MC simulations (dotted line),  JK simulations (solid line with errors) and JK for real LRG data (big dots) 
	 \label{fig:errperp}}
\end{figure}

We calculate the projected correlation function $\Xi(\sigma)$ for each group mock
integrating through $\pi$ the redshift space correlation function $\xisp$, and
we also calculate the JK error. Then we look at the difference between the MC
dispersion and the mean over all the JK errors. In Fig.\ref{fig:errperp}, top
panel, we see the mean $\Xi(\sigma)$ over all the simulations (solid line with
errors), and over-plotted in blue big dots the $\Xi(\sigma)$ for the real LRG data. In
the bottom panel we show a comparison
of the different errors in the simulations
(MC dotted line, and JK solid line with errors) and the error JK in the LRG data (big dots).
Errors in MC and JK in simulations coincide at the scales where we can
use the $\Xi(\sigma)$ (below 30Mpc/h). The JK error in LRG is also
similar to the errors in the simulations.
We have done the same analysis with dark matter simulations and they also work well, as expected. We conclude that using JK errors gives a good approximation to the true MC errors.

\subsection{Errors in the real-space correlation $\xi(r)$}
\label{sec:errorsreal}

\begin{figure}
	\centering{ \epsfysize=6cm\epsfbox{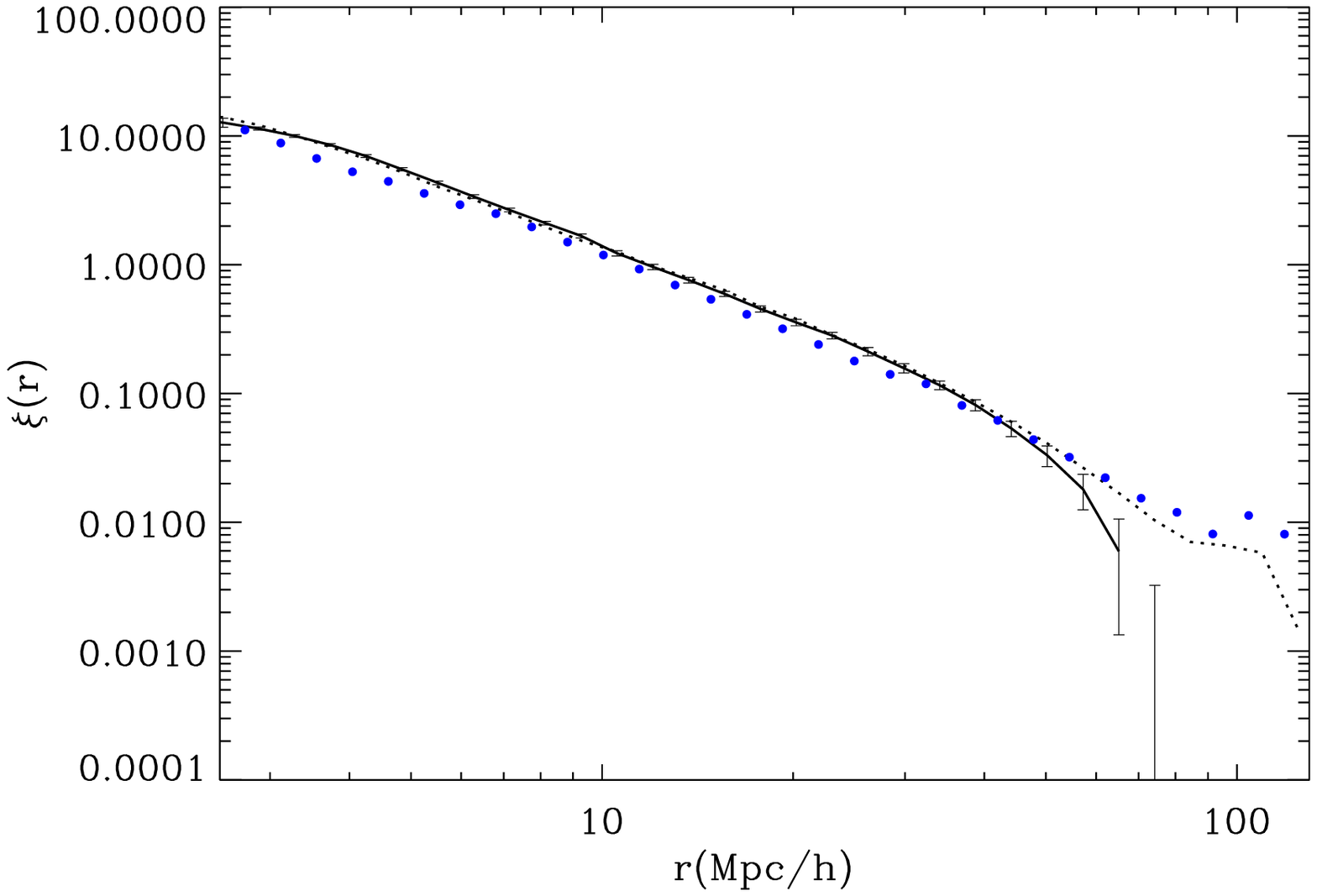}}
	\centering{ \epsfysize=6cm\epsfbox{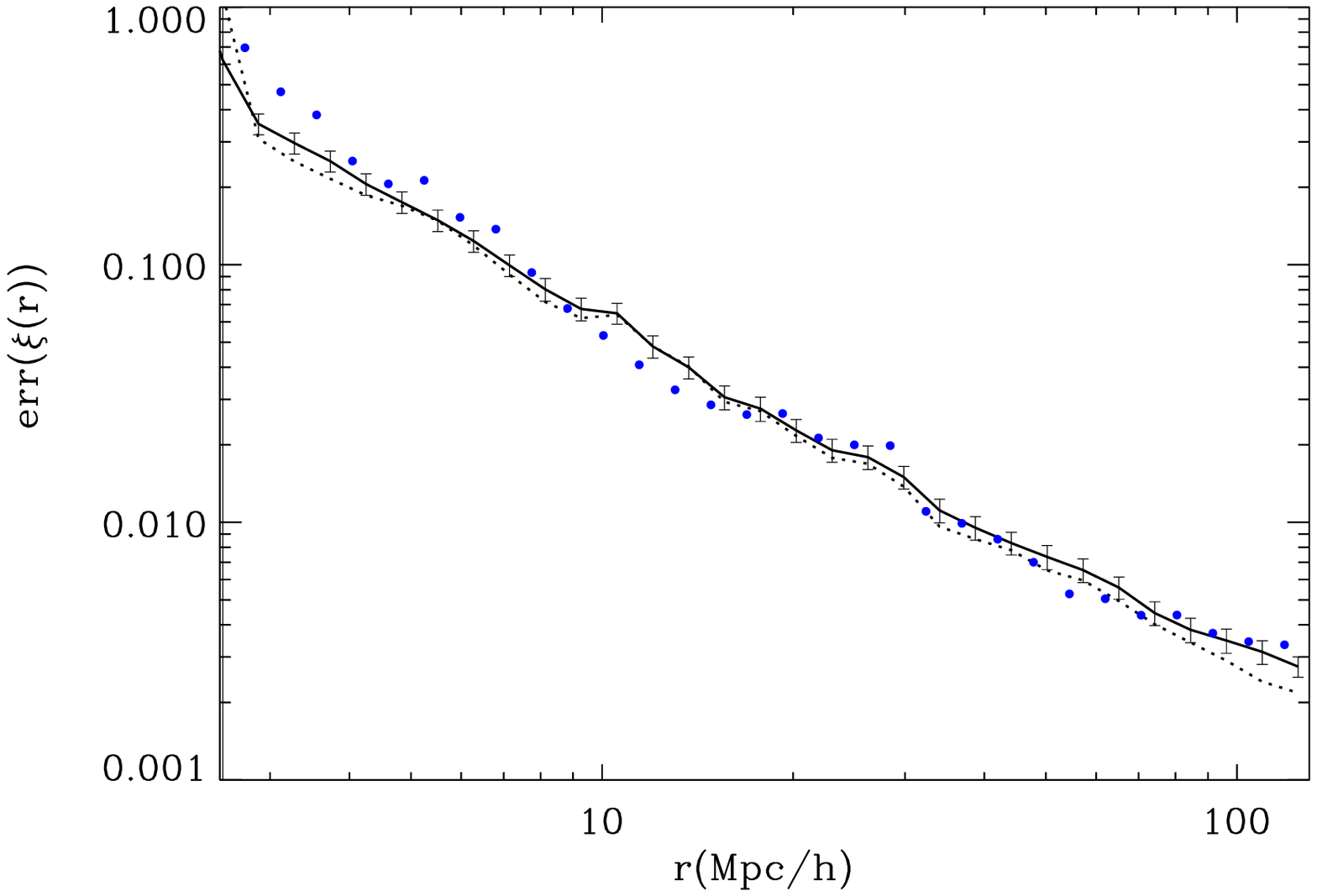}}
	\caption{Top: $\xi(r)$ for the group simulations calculated from 
	$\Xi(r)$ (solid line with errors) and the value for LRG data (big dots).
	 Bottom: Errors in $\xi(r)$ for MC simulations (dotted line),  JK simulations (solid line with errors) and JK for real LRG data (big dots) 
         \label{fig:errrealsp}}
\end{figure}

We now calculate the real-space correlation function $\xi(r)$ from the projected
correlation function $\Xi(\sigma)$ (Eq.(\ref{eq:xirr2})). In Fig.\ref{fig:errrealsp},
top panel, we see the correlation function obtained from deprojecting $\Xi(\sigma)$
with errors (solid line) compared to the real-space correlation function which we can
calculate using the simulations in real space, without redshift distortions
(dotted line). We can see in Fig.\ref{fig:errrealsp} how we recover well 
$\xir$ for small scales (below 40Mpc/h). Over-plotted
in blue (big dots), we see the real-space correlation function obtained in LRG data, 
with a similar 
bias than the group simulations, as expected ($b$ was found to be $b\simeq 2$
in Paper I). On the bottom panel of Fig.\ref{fig:errrealsp}, the plot 
compares errors in MC (solid line with errors) and JK (dotted line) case, and they are 
very similar at small scales. We have over-plotted the error JK in LRG data (big dots).
We have also done a similar analysis with dark matter
mocks and they also show a good agreement between JK
and MC errors.
As in the previous case, we conclude that using JK errors give a good approximation to the true MC errors.


\subsection{Validity of the models}\label{sec:validity}

\begin{figure}
	\centering{\epsfysize=8cm\epsfbox{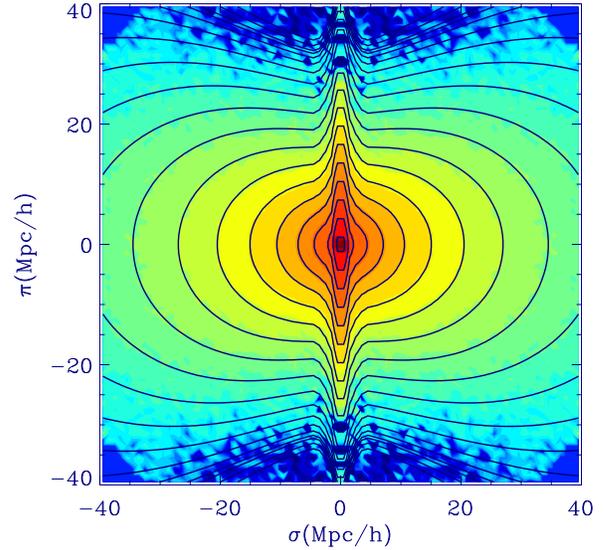}}
	\caption{$\xisp$ for the MICE dark matter simulation (as colors). The contour colors are -0.05, -0.01, -0.005, -0.001, 0, and 0.001 to 20 with 20 equally spaced logarithmic bins. In solid lines we have plotted the model for the input parameters of the simulation. For $\sigma$ larger than 5Mpc/h we use the effective $\sigma_{12}=400km/s$ (asymptotic value at large scales) to model $\xisp$ while for $\sigma$ lower than 5Mpc/h we use a different $\sigma_{12}$ as seen in Fig.\ref{fig:pairwise7680}, for each $\sigma$ constant along the LOS $\pi$ \label{fig:corrmice}}
\end{figure}

\begin{figure}
	\centering{ \epsfysize=6cm\epsfbox{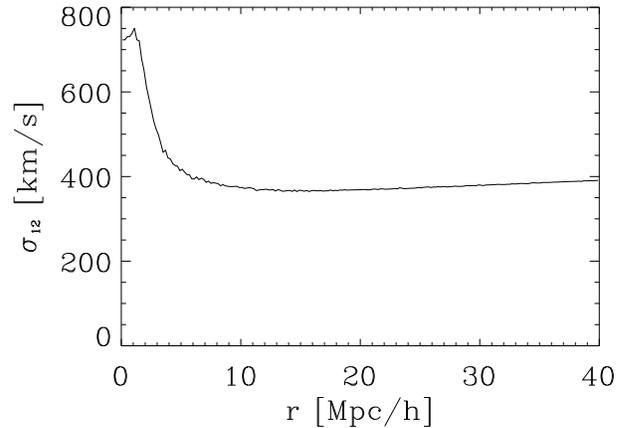}}	
	\caption{Dispersion in the pairwise velocity distribution $\sigma_{12}$
estimated from the velocity field 
in the MICE dark matter simulations as a function of the distance between particles.
\label{fig:pairwise7680}}
\end{figure}

Besides studying errors, we also use the simulations to test the methods that
we will apply to real data (LRG). In top panel
of Fig.\ref{fig:errrealsp} we showed
that we can recover the real-space correlation function. Now we want to see if
we can model the redshift-space correlation function with the simple 
model explained in \S\ref{sec:model}, when taking the approximation of $\sigma_{12}$ constant along the LOS and varying it for $\sigma$ lower than 5Mpc/h. 
In Fig.\ref{fig:corrmice} we plot $\xisp$ at
small scales for the mean of the MICE dark matter mocks. We  over-plot in solid lines 
our model with a varying  pairwise velocity dispersion. We based this model
on the pairwise velocity dispersion $\sigma_{12}(r)$ estimated from the velocities
in the simulations. Fig.\ref{fig:pairwise7680} shows the measured pairwise 
velocity dispersion $\sigma_{12}(r)$. In our model, for $\sigma$ larger than 5Mpc/h 
we use the effective $\sigma_{12}=400km/s$
(asymptotic value at large scales). For $\sigma$ lower
than 5Mpc/h  we use a different $\sigma_{12}$ as given by  
Fig.\ref{fig:pairwise7680}. With this simple approximation, we 
reproduce very well the observed FOG as shown in Fig.\ref{fig:corrmice}.
We have done this study with dark matter mocks since we know the variation of $\sigma_{12}$ exactly, while when we construct the group mocks, 
we loose the pairwise velocity at small scales due to the way we identify groups.


\section{The Data: Luminous Red Galaxies}    
\label{sec:data}

SDSS luminous red galaxies (LRGs) are selected on the basis of color and
magnitude to have a sample of luminous intrinsically red galaxies that extends
fainter and farther than the SDSS main galaxy sample. 
Eisenstein et al (2001) give an accurate description of the sample. In Paper I we give more details of
our selection. Here we just give a summary.

We k-correct the r magnitude using the Blanton program 'kcorrect'
\footnote{http://cosmo.nyu.edu/blanton/kcorrect/kcorrect\_help.html}. We need to
k-correct the magnitudes in order to obtain the absolute magnitudes and
eliminate the brightest and dimmest galaxies. We have seen that the previous
cuts limit the intrinsic luminosity to a range $-23.2<M_r<-21.2$, and we only
eliminate from the catalog some few galaxies that lay out of the limits. Once we
have eliminated these extreme galaxies, we still do not have a volume limited
for high redshift galaxies, but we suppose that the variations in luminosity
just change the overall shape in the clustering.

We have masked the catalog using at the first step the photometric DR6 mask,
based on the number of galaxies per pixel. In previous works we saw that the
mask that we obtain statistically by dropping out the pixels with small number
of galaxies gives the same correlation function that the one obtained by
extracting the polygons masked by the SDSS team. After that, we compare our
masked catalog to the LRG spectroscopic catalog, and we extract the galaxies
that lay outside from ``good'' plates.

This rough mask could imprint spurious effects at very small scales, but we are
not interested in these scales where fiber collisions in the redshift catalog
are limiting our analysis, for distances less than 55arc sec, less than 0.3Mpc/h
at the mean redshift of LRG data, z=0.35. We obtain 75,000 galaxies for the
final catalog, from z=0.15 to z=0.47. The area of the data used is around 13.5\%
of the sky. See Fig.2 in 
Cabr\'e and Gazta\~{n}aga (2008) for a plot of the mask. Recently we have used another mask, provided by
Swanson et al. (2008), which is in a readily
usable form, translating the original mask files extracted
from the NYU Value-Added Galaxy Catalog (Blanton et al.
2005), from MANGLE into Healpix format (Gorski et al.
1999), and results are very similar.
We have also done the analysis with the recent data release DR7, the final one in SDSS, and results are very similar at small scales.
   
We use the $\xi$ estimator of Landy and Szalay (1993),

\begin{equation}
\xisp  = \frac{DD - 2DR + RR}{RR}
\end{equation}

to estimate the 2-point correlation function in redshift space, with a random catalog 20 times denser than the SDSS catalog. The random catalog has the same redshift distribution as the data, but smoothed to avoid the elimination of intrinsic correlations in the data, and also the same mask. We count the pairs in bins of separation along the line-of-sight (LOS), $\pi$,  and across the sky, $\sigma$. The LOS distance is just the difference between the comoving distances in the pair. The perpendicular distance is $\sigma=\sqrt{s^2-\pi^2}$, which corresponds approximately to the mean redshift. 

\section{Analysis}

As we explore small scales, we will show in this section  that we encounter
 the following problems. First of all, the
bias becomes clearly dependent on the scale for distances smaller than
about 10 Mpc/h, because LRG are galaxies highly biased so they only keep the linear
bias constant at large scales. Secondly, the model that we are using 
assumes that the pairwise velocity dispersion is independent of scale, which
is not a good approximation for small scales as we have seen in the simulations.
We can arrange this by using different $\sigma_{12}$ for each real 
distance $r$, as explained in Section \ref{sec:model}. 
Here we want to check if we can infer the correct variation of $\sigma_{12}(r)$
with scale using the $\xisp$ data. As a test, we will compare the inferred values of 
$\sigma_{12}(r)$ with the  direct measurement from the
velocity field in the simulation.

\subsection{Real space correlation}

\begin{figure}
\centering{ \epsfysize=6cm\epsfbox{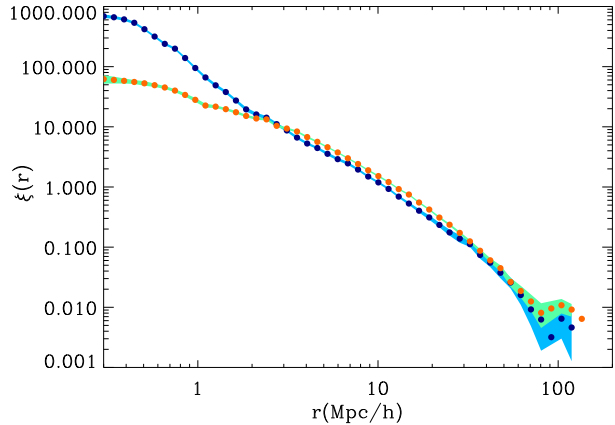}}
\caption{$\xi(r)$ (blue dots) and $\xi(s)$ (orange dots). We see clearly 
how the redshift-space correlation function is the real-space 
correlation function biased by a constant factor that represents 
gravitational infall (dependent on $\beta$ in Kaiser approximation) 
at large scales above 4Mpc/h (although $\xir$ is overestimated for 
scales larger than 30Mpc/h due to the precision in the calculation). 
However, for small scales, the redshift-space $\xis$ is strongly 
suppressed compared to $\xir$ due to random peculiar velocities. 
\label{fig:corrplot}}
\end{figure}

In Fig.\ref{fig:corrplot} we show the resulting real-space correlation function
which we have calculated using Eq.(\ref{eq:xirr2}) (in blue) and over-plotted
the monopole in redshift space (in orange). At intermediate scales, from 5 to 30
Mpc/h (the top value limited by the method to obtain $\xir$), $\xis$ is equal to
$\xir$ but biased by a constant factor, as in Eq.(\ref{eq:xisr}), due to Kaiser
effect. We had shown this constant bias at intermediate scales (around 10Mpc/h) in
Fig.8 of paper I \cite{paper1}, where we plot the ratio of the correlation function in
redshift space and real space $\xis/\xir$. At large scales, 
we can associate it to a function
of $\beta$ obtained in Paper I at large scales. The agreement
is excellent, which provides a good consistency check for our results. The
difference between the real and redshift space correlation function at small
scales is primordially due to the random peculiar velocities.

\begin{figure}
\centering{ \epsfysize=6cm\epsfbox{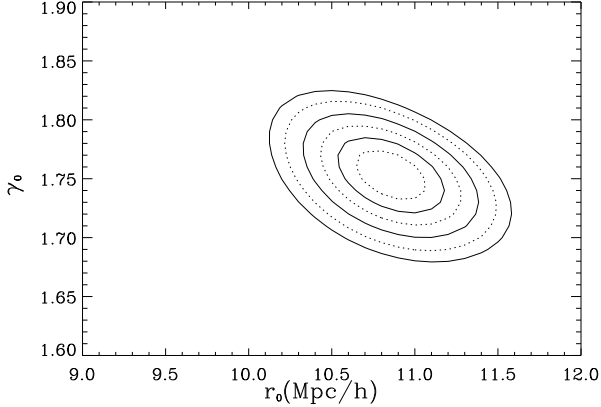}}
\caption{Best fit of $\gamma_0$ and $r_0$
of $\xi(r)$ to a power law model $\xi(r)=(r/r_0)^{-\gamma_0}$,
1-15Mpc/h. Dotted and continuous contours correspond to 1 ad 2 degrees of freedom 
with 1,2 and 3-sigma confidence.
\label{fig:corrfit1}}
\end{figure}

\subsection{Power law fit}

\begin{figure}

\centering{ \epsfysize=6cm\epsfbox{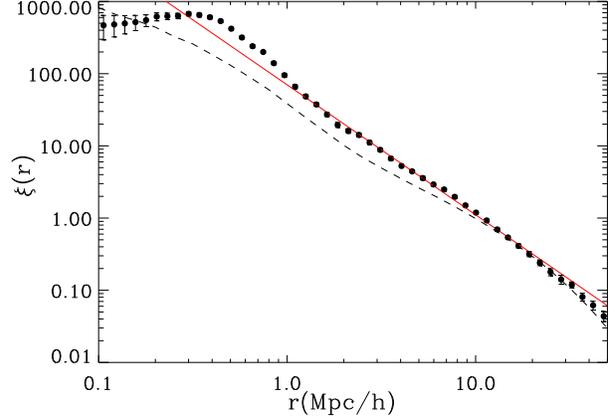}}
\caption{Observed $\xir$ (symbols with
errorbars), best fit to the power law
$\xi(r)=(r/r_0)^{-\gamma_0}$ (red), and dashed over-plotted the best real space
correlation function for large scales, assuming a constant bias
\label{fig:corrfit2} }
\end{figure}

We next fit our estimation of
the real space correlation function to a power law,
$\xi(r)=(r/r_0)^{-\gamma_0}$, from 1Mpc/h to 15Mpc/h. At scales smaller than
1Mpc/h the fit is not good, it is no longer a power law. In Fig.\ref{fig:corrfit1} we
show a fit for $r_0$ and $\gamma_0$. In Fig.\ref{fig:corrfit2} we show the measured
real space correlation function, the best fit power law model (red) 
and the
dark matter non-linear correlation function obtained from best  parameters in
Paper I 
 which works well for a linear bias, ie
on large scales. The large scale and small
scale fittings models agree well at large scales,
as can be seen in the plot, and the correlation function
does not follow a power law for distances smaller than $\backsim$ 1Mpc/h, where
we can probably see the transition from the one-halo to the two-halo term
\cite{scoccimarro}.

These results are similar to other studies.
Zehavi et al (2005) did a similar analysis with a previous SDSS spectroscopic
data release (35000 LRGs) at intermediate scales from 0.3 to 40Mpc/h. We have
doubled the number of LRGs and our results agree with them for the monopole, the
projected correlation function, and the obtained real-space correlation
function, with the same main conclusions. Also 
Eisenstein et al (2005) , in a study of
small scales (0.2-7Mpc/h) using the cross-correlation between spectroscopic LRG
with the main photometric sample, remark that $\xir$ can not be explained with a
power-law fitting. However, 
Masjedi et al (2006) have obtained the correlation
function at very small scales (0.01-8Mpc/h) and have found that, although with
some features diverging from a power law,  all the range is really close to a
$\xir\propto r^{-2}$.

\subsection{Non-linear bias}

We have checked in our simulations with halos that the non-linear bias 
typically follows a
power-law (at least for distances larger than 0.3Mpc/h), which has a different slope $\gamma_b$ depending on the halo mass. 
In general this could depend on
 other parameters concerning galaxy formation.
LRG are assumed to be red galaxies that trace halos of $10^{13}M_\odot$, but
there is a wide range of halos masses, and the non-linear bias shows us these
properties. Here we estimate the bias as

\begin{equation}\label{eq:bnolineal}
b(r)=\sqrt{\xi(r)/\xi(r)_{DM}}=b\;b_{nl}(r)
\end{equation}

where we have separated the bias $b(r)$ into a constant scale independent
term, $b$, on large scales and a function of scale $b_{nl}(r)$ which goes
to unity on large scales.
When $b_{nl}(r)$ reaches unity, we assume that the bias is linear from
there on. We define a parameter $r_b$ for the power law  
$b_{nl}(r)=(r/r_b)^{-\gamma_b}$ which shows the scale
from which non-linearity begins to be important. The value of
$r_b$ should coincide approximately
with the correlation length, where the real space correlation function is 
unity. 

\begin{figure}
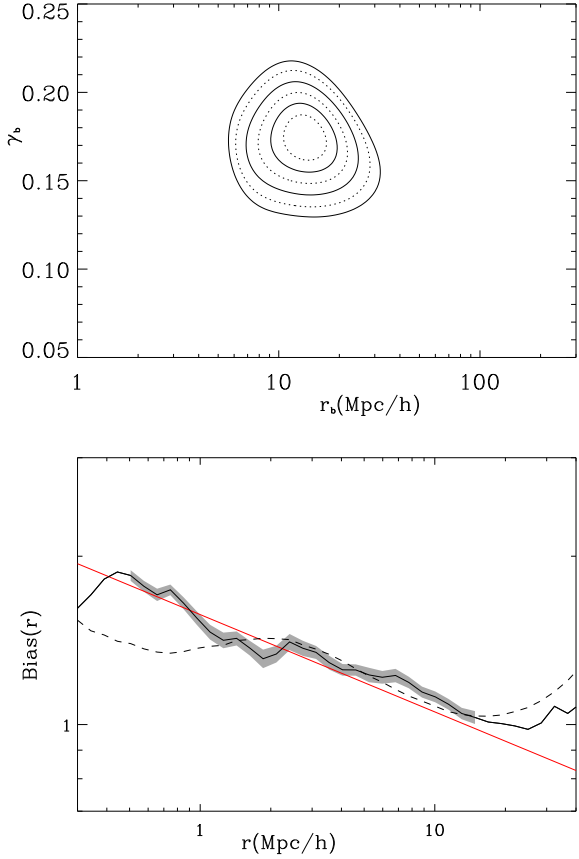


\centering{ \epsfysize=6cm\epsfbox{figures/fitcorrbiasnl11.0.5.15.ps1}}
\centering{ \epsfysize=6cm\epsfbox{figures/fitcorrbiasnl11.0.5.15.ps2}}
\caption{Top panel: Best fit to no linear bias $b_{nl}(r)$ (defined in the text)
with a power law $b_{nl}(r)=(r/r_b)^{-\gamma_b}$. Bottom panel: Non linear bias
$b_{nl}(r)$ (solid line with errors in gray) and best power law fit (red). We
have also over-plotted in dashed line the bias obtained if we suppose that the
galaxy correlation function is a power law. \label{fig:biasfit}}
\end{figure}

To compute the non-linear bias $b(r)$ from the above equation,
we need the dark matter correlation function and the
linear bias $b$, which we have taken from the fitting at large scales
done in Paper I. We have
calculated the bias for all the values of $\Omega_m$ and 
different amplitudes. Then we marginalize
over them. In Fig.\ref{fig:biasfit} we see the contours for $r_b$ and $\gamma_b$
(top panel), and the best fit (red in the bottom panel). This fit to the bias
can explain the differences seen previously between  the correlation function
and a power law for scales smaller than 1Mpc/h. At scales smaller than 0.3Mpc/h,
the real space correlation function turns down due to fiber collisions \cite{masjedi}.
In detail, we see in Fig.\ref{fig:biasfit} a feature in the bias between 1 and 2
Mpc/h, indicating that LRGs galaxy bias is not completely smooth. 
We think that this feature is due to the range
of halo sizes of our LRGs, which makes it difficult to predict exactly the
transition point from the 1-halo to the 2-halo term \cite{scoccimarro}.
If galaxies are residing within dark matter halos then the clustering of the
galaxies on scales larger than halos is determined by the clustering of the dark
matter halos that host them, plus statistics of the occupation of halos by
galaxies. For larger scales than $\simeq$ 2Mpc/h (the biggest halos), the
clustering comes entirely from LRGs that reside in different halos, while for
smaller scales, the clustering can come from galaxies in different halos or
galaxies in the same halo until it is reached a minimum size of halos (see
Masjedi et al (2008) for a more detailed explanation).

\subsection{Monopole and Quadrupole}\label{sec:changesv}

\begin{figure}
\centering{ \epsfysize=6cm\epsfbox{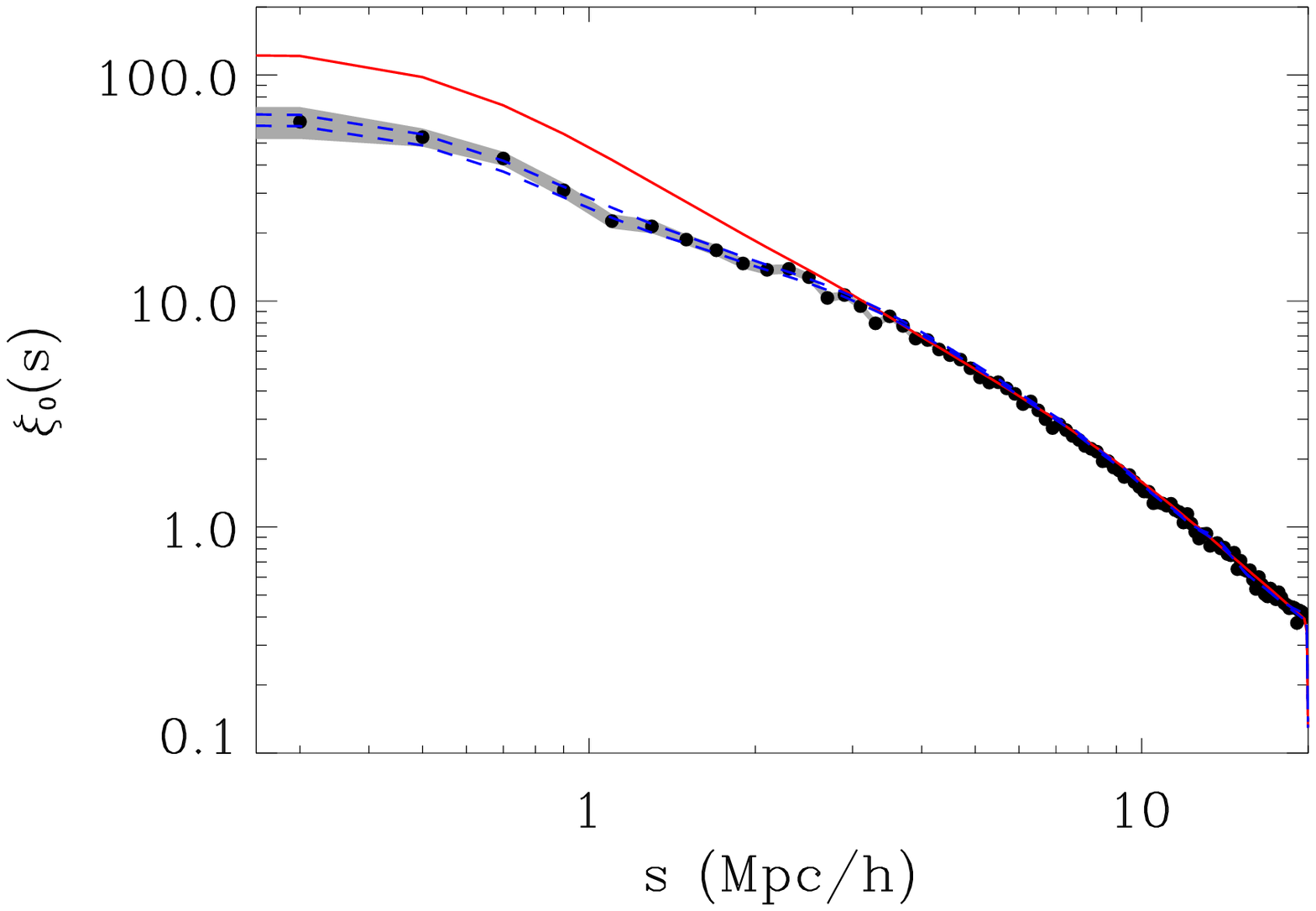}}
 \centering{ \epsfysize=6cm\epsfbox{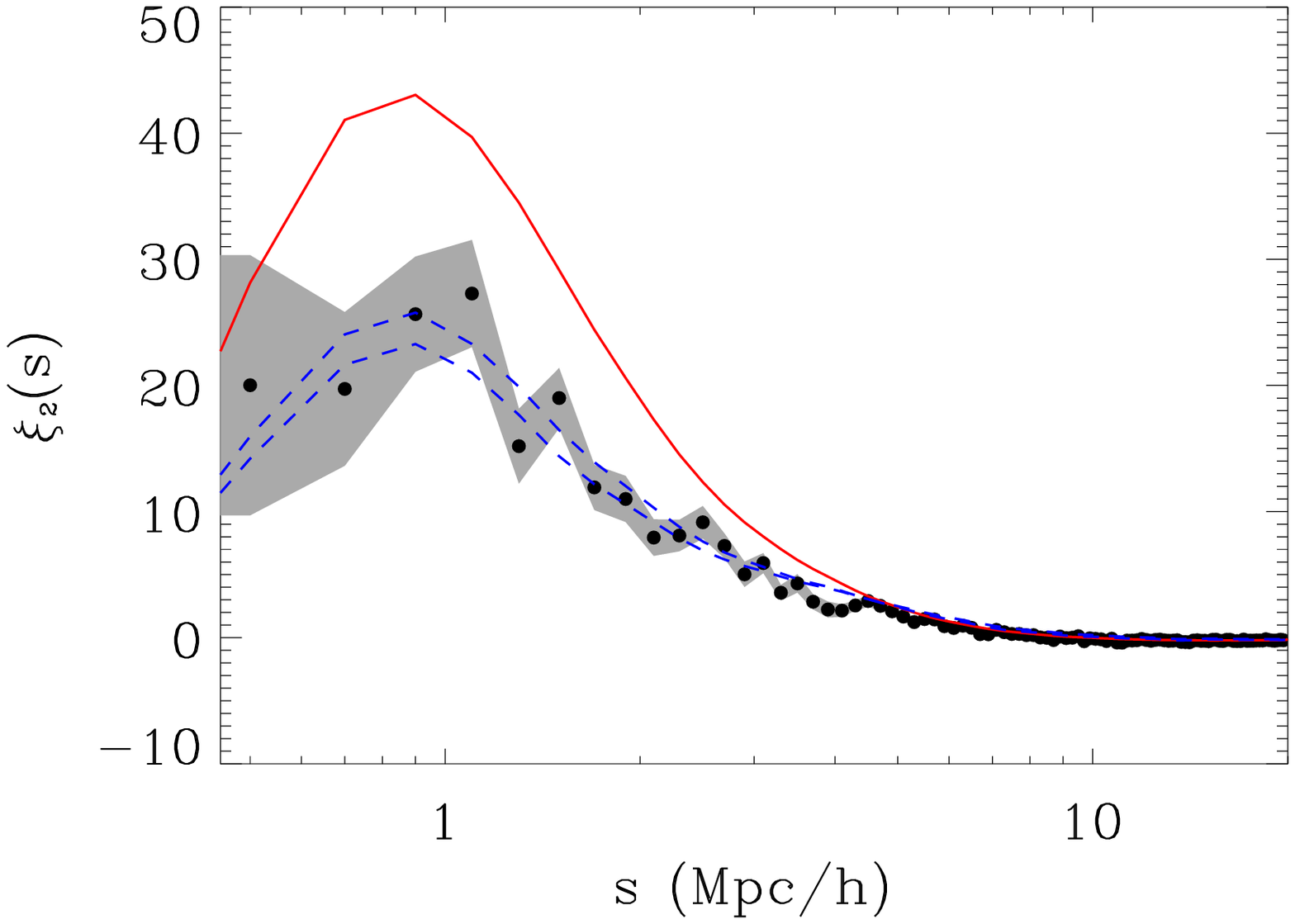}}
\caption{Monopole $\xis$ (dots) with errors (gray), best model assuming a
constant $\sigma_{12}$ (solid red) and model
assuming variation in $\sigma_{12}$ with errors (dashed blue), as in Fig.\ref{fig:realsigmavchange}. Bottom panel: Quadrupole $\xi_2(s)$ (dots) as in the top panel.
\label{fig:monopole1Mpc}}
\end{figure}

Once we have obtained the real space correlation function, we can look at the
monopole $\xis$, the quadrupole $\xi_2(s)$ and also at $\xisp$ in order to check
the result. We see in Fig.\ref{fig:monopole1Mpc} the monopole $\xis$ (top panel)
and quadrupole $\xi_2(s)$ (bottom panel) binning the distance with 0.2Mpc/h, and
over-plotted in red solid line the theoretical model, which we have found integrating
$\xisp$ and assuming a constant $\sigma_{12}$ (derived using the
normalized quadrupole Q(s) in Paper I). The prediction uses the model
explained above and in Paper I, where the shape of $\xisp$ is given by
the real-space correlation function and the parameters describing the 
velocity distortions. The monopole directly measured in the
data (dots with shaded region in top panel of Fig.\ref{fig:monopole1Mpc})
does not agree with the model (red line), which is lower for scales smaller than 3Mpc/h.
The same happens to the quadrupole (bottom panel), where the model is also higher than
the measurements. These differences indicate that $\sigma_{12}$ is higher at 
smaller scales as shown in Fig.\ref{fig:pairwise7680} for simulations.

\subsection{Recovering $\sigma_{12}$ from simulations}\label{sec:changesv1}

We now recover the dispersion of pairwise velocities $\sigma_{12}$ at each perpendicular distance $\sigma$ 
assuming that $\sigma_{12}$ remains constant along the LOS $\pi$. In reality, $\sigma_{12}$
 is a function of the real distance r, $\sigma_{12}(r)$. At each $\pi$, $\sigma$, the correlation is a convolution of all the real distances. We could in principle
 try to fit all the $\pi$-$\sigma$ plane for different shapes of $\sigma_{12}(r)$, but this
involves too much freedom and such a fit is not feasible in practice. 
The alternatives are to do the fit in Fourier space, where there is no dependence 
of $\sigma_{12}(k)$ along the LOS, or to use the approximation 
$\sigma_{12}(r)\sim\sigma_{12}(\sigma)$ in the zone of the $\pi-\sigma$ plane where this 
is a good approximation. 
We have studied the differences between $\xisp$ obtained from either
$\sigma_{12}(r)$ or from $\sigma_{12}(\sigma)$ to explore this later possibility. 
The two models differ, more or less strongly depending on the case, 
at small $\sigma$ and large $\pi$. For our dark matter simulations, the differences 
are smaller than in the LRG case, where they can be large enough to bias the final result.
 The best option, then, is to fit for each perpendicular distance 
$\sigma_{12}(\sigma)$ using $\xisp$ up to a maximum $\pi_{max}=5-10Mpc/h$, where both models
agree well. If we go further in $\pi$, the result is biased to lower values of $\sigma_{12}$.

\begin{figure}
\centering{ \epsfysize=6cm\epsfbox{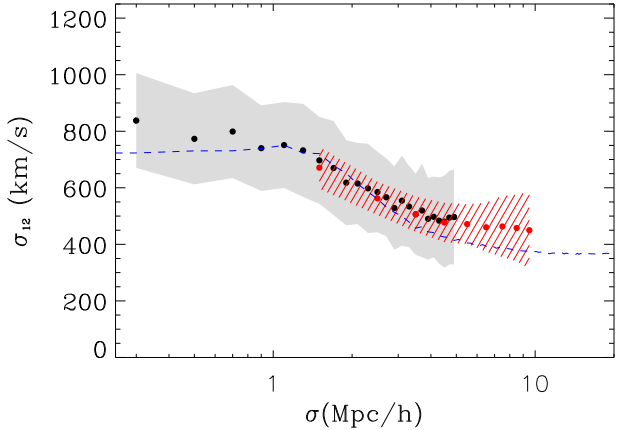}}
\caption{We plot the mean recovered $\sigma_{12}(r)$  ($\pi_{max}=5Mpc/h$) for a bin
 0.2Mpc/h (black circles with gray zone as errors) and for 1Mpc/h (red circles
 with dashed zone errors), compared to the original $\sigma_{12}(r)$ (dashed line)
 calculated in the real space simulation. We take each mock redshift space 
$\xisp$ and model it following Eq.(\ref{eq:hamiltonmethod}), from where we 
obtain the dependence of $\sigma_{12}$ on the distance.
\label{fig:changesvsim}}
\end{figure}

\begin{figure*}
\centering{ \epsfysize=3cm\epsfbox{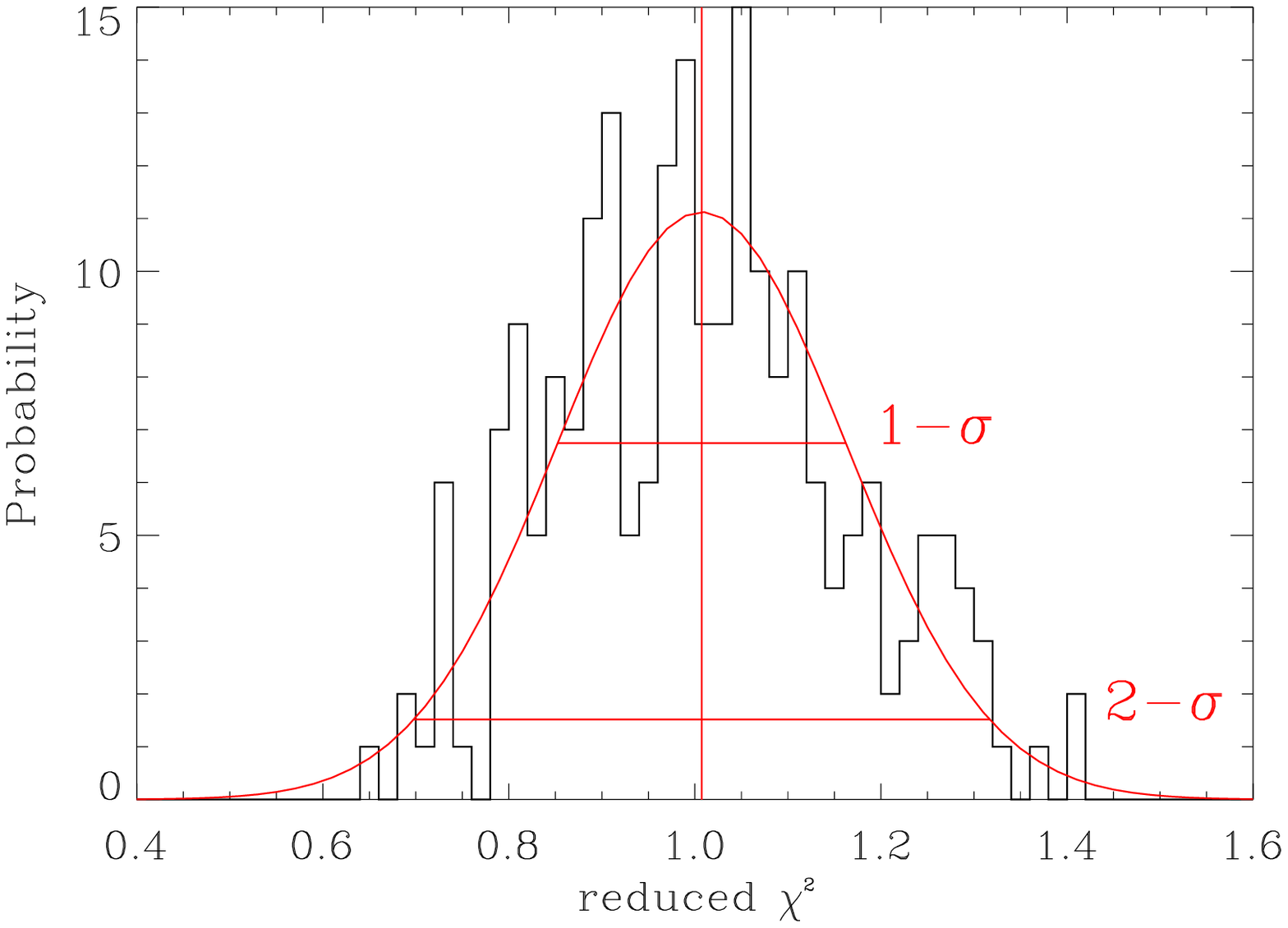}}
\centering{ \epsfysize=3cm\epsfbox{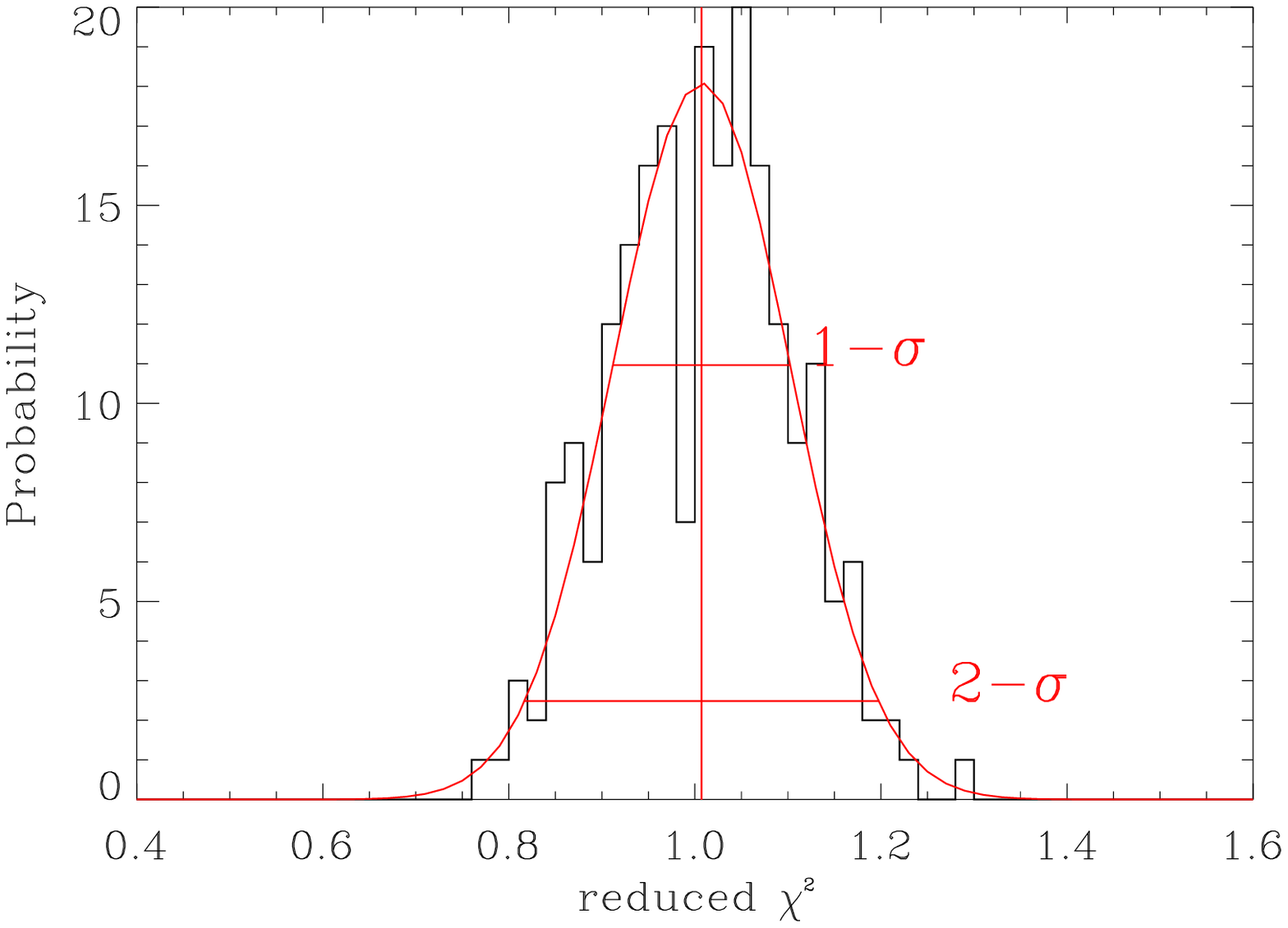}}
\centering{ \epsfysize=3cm\epsfbox{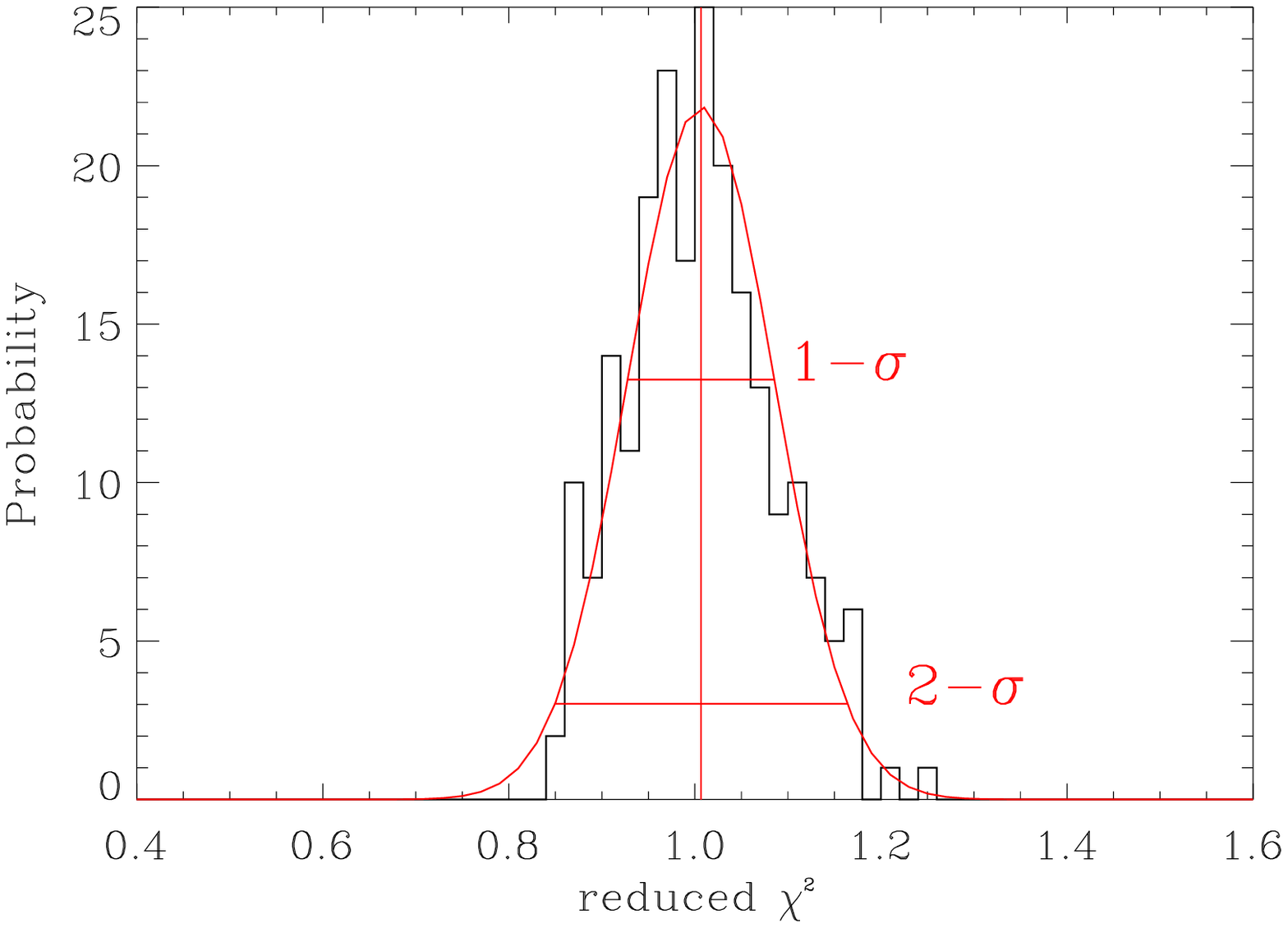}}
\centering{ \epsfysize=3cm\epsfbox{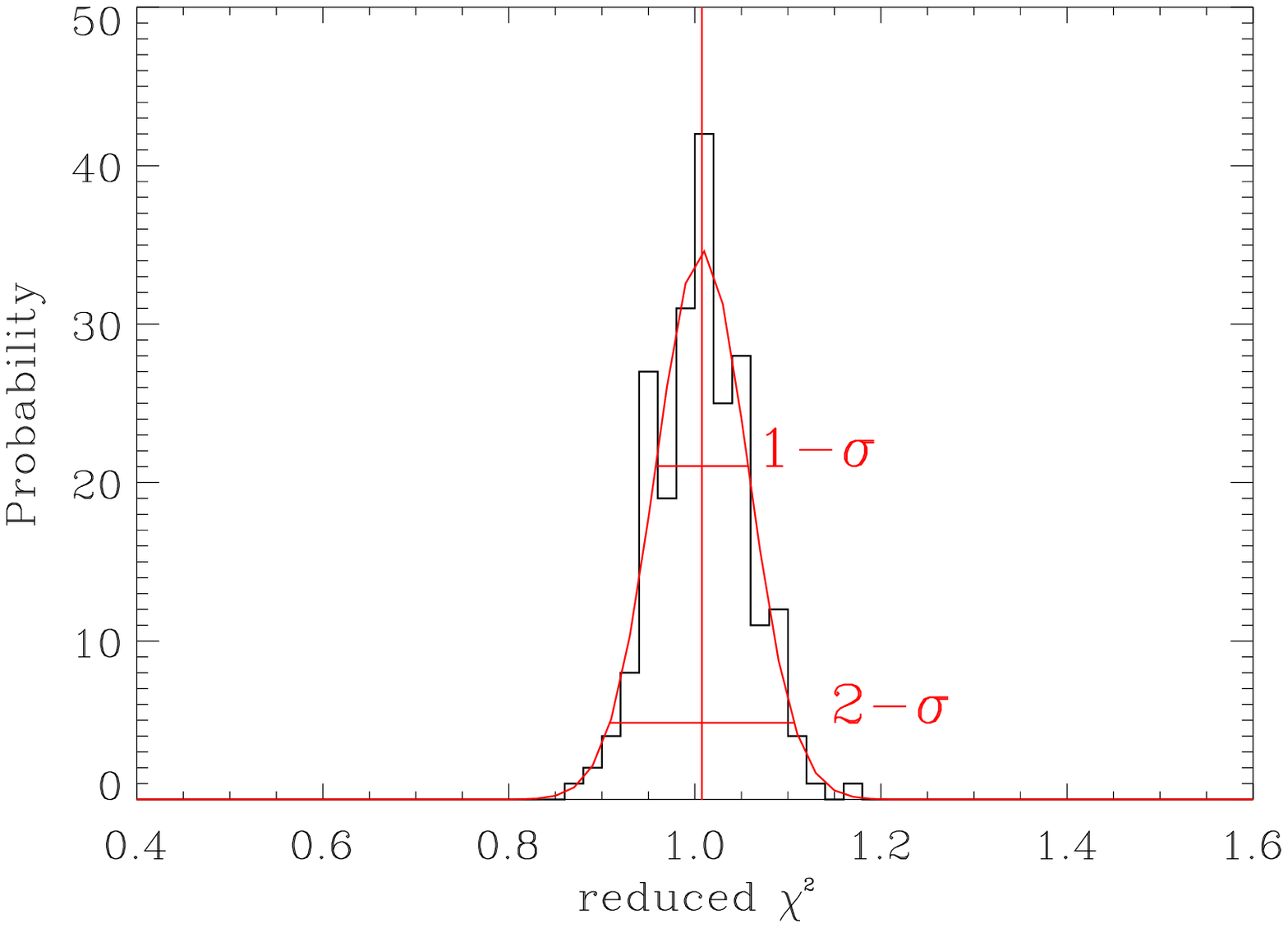}}
\caption{Distribution of $\chi^2$ obtained from the mocks when we fit each $\xisp$ with the dark matter model which includes the correct $\sigma_{12}(r)$. Different panels are showing the zone where we fit, from smaller to larger, for $\xisp > 1., 0.57, 0.43, 0.24$
\label{fig:reducedchi2}}
\end{figure*}

In order to test the method to recover $\sigma_{12}(r)$ we have used our dark matter mocks, 
where we trust velocities to be accurate. 
We take each mock and recover $\sigma_{12}(r)$ for $\pi_{max}=5, 10, 18, 30 Mpc/h$. 
We have used a bin of 0.2Mpc/h for $\sigma$ = 0.3 - 5Mpc/h and a bin of 1Mpc/h for $\sigma =$ 1 - 10Mpc/h.
 In Fig.\ref{fig:changesvsim} we plot the mean recovered $\sigma_{12}(\sigma)$ for 0.2Mpc/h 
(black circles with gray zone as errors) and for 1Mpc/h (red circles with dashed zone errors), 
compared to the original $\sigma_{12}(r)$ calculated directly from velocities in the
simulation (dashed line). 
We use $\pi_{max}=5Mpc/h$ for this plot, but we obtain very similar results when we go further in $\pi$. 
The main conclusion of this analysis is that we recover correctly the random 
dispersion of pairwise velocities, and secondly that the method to derive $\sigma_{12}$
 works at least until $\pi_{max}=30Mpc/h$, for dark matter. This method is good to obtain
 the variation of $\sigma_{12}$ at small scales. At larger scales (around 5Mpc/h), 
$\sigma_{12}$ reaches a constant, and we see that in our study the estimated $\sigma_{12}$ 
is slightly biased to higher values. At these distances, variations on $\sigma_{12}$ 
do not change much the model for $\xisp$, so this is not a problem for comparison 
to data. We should take into account that we are using a simplified model which 
assumes that pairwise velocities are exponential. This is a good approximation 
for small scales, as we see from the simulations, but the real dispersion is in fact skewed and
 it is not perfectly exponential, so we can see slight variations from the real dispersion.
However, to recover an unbiased result for $\sigma_{12}$ from $\xisp$, 
the most important point is to take into account the infall velocities 
in the model (Scoccimarro 2004). Here we include infall velocities through the parameter $\beta$.

We next calculate the $\chi^2$ for each mock in the central region of the plane $\pi-\sigma$ (defined
buy the contours of $\xisp > 1.0, 0.57, 0.43, 0.24$) and we obtain a mean and dispersion of 
the reduced $\chi^2$ (ie the $\chi^2$ divided by the number of degrees of freedom). 
The distribution of reduced $\chi^2$ values from the different mocks
can be seen in Fig.\ref{fig:reducedchi2}, the mean is equal 1 and this means that the 
model works perfectly well for our simulations of dark matter. But note how the dispersion 
in the distribution of $\chi^2$ is quite broad, there is therefore room for some divergences away 
from 1.

\subsection{Measurement of $\sigma_{12}$ as a function of scale}\label{sec:changesv2}

\begin{figure}
\centering{ \epsfysize=6cm\epsfbox{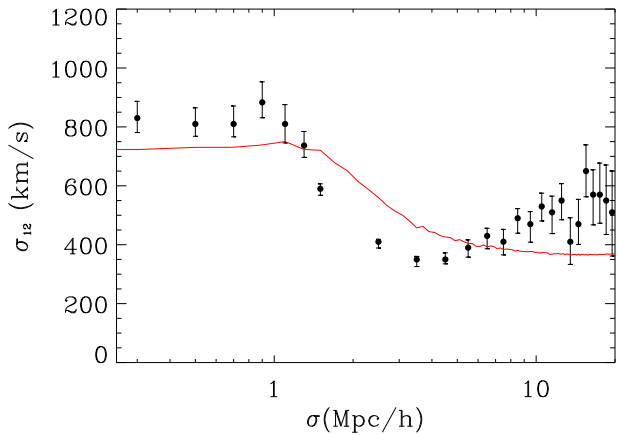}}
\caption{$\sigma_{12}$ vs distance for LRG galaxies (circles with errorbars), 
obtained from modeling $\xisp$ with Eq.(\ref{eq:hamiltonmethod}). 
We compare it to the true value from velocities in the
dark matter simulation (solid line).
\label{fig:realsigmavchange}}
\end{figure}

Once we have showed that we can recover $\sigma_{12}(r)$ and that this simple model works 
for dark matter, let's do the same for the LRG galaxies. In order to calculate the model 
$\xisp$ at each $\sigma$ and $\pi$, we need to assume the distortion parameter 
$\beta$ and the real space correlation function $\xir$. 
We use $\beta=0.34$, found in Paper I of this series. 
We also use the real space correlation function $\xi(r)$ from the integration 
of $\xisp$ along the line-of-sight, see Eq.(\ref{eq:xirr}) and Fig.\ref{fig:corrplot}. 
As mentioned above, the real space correlation
$\xi(r)$ is not well recovered above 30Mpc/h, where we will use instead the measured monopole 
corrected by the distortion bias factor, Eq.(\ref{eq:xisr}). 
At large scales, the real space and the redshift space correlation function are almost 
linearly biased, except from the BAO peak where there are some smaller
non-linear effects again, but these do not have an effect in the modelization
 of $\xisp$ at small scales, so this is a good approximation.
 We calculate $\sigma_{12}(\sigma)$ up to $\pi_{max}=5, 10, 15, 20, 25 Mpc/h$. 
For values above $\pi_{max}=$10Mpc/h, we find that
$\sigma_{12}(\sigma)$ starts to be biased low for small values of $\sigma$ with respect to the
true values of $\sigma_{12}(r)$.
For higher values of $\sigma$, we can fit until a larger $\pi_{max}$, reducing the errors. 
We use the same binning as in the simulations and the resulting $\sigma_{12}(r)$ 
is plotted in Fig.\ref{fig:realsigmavchange}, where we compare it to the true value from velocities
in simulations (solid line). Both results have approximately the same amplitude, 
that depends strongly on the cosmological parameters. If there is no velocity bias, then 
our universe must be similar to our simulations.

Now we can recalculate the monopole and quadrupole including the variation of $\sigma_{12}$ with
scale. We obtain a good fit to the results, as we can see in Fig.\ref{fig:monopole1Mpc}
( as a blue dashed line) compared to the same prediction assuming a constant $\sigma_{12}$ 
(solid red line).

 As mentioned in \S\ref{sec:intro},
Slosar et al (2006) show that pairwise velocity distribution in real space is a
complicated mixture of host-satellite, satellite-satellite and two-halo pairs.
The peak value is reached at around 1Mpc/h and does not reflect the velocity
dispersion of  a typical halo hosting these galaxies, but is instead dominated
by the sat-sat pairs in high-mass clusters. 
Tinker et al (2007) uses the HOD model
to explain that at r $\backsim$ 1 - 3 Mpc/h, the PVD rapidly increases towards smaller scales as
satellite-satellite pairs from massive halos dominate. At r $<$ 1 Mpc/h, the
pairwise velocities dispersion decreases with smaller separation because
central-satellite pairs become more common, but we do not see this tendency because we do not study such small scales due to fiber collisions. At r $>$ 3 Mpc/h, the $\sigma_{12}$ is dominated by two-halo central galaxy pairs. Their predictions agree well with
our results in shape and amplitude (compare Fig.\ref{fig:realsigmavchange} with their Fig.4 more luminous prediction). Note that the amplitude of the predictions can be increased or reduced easily by changing the 
cosmological parameters.

Tinker et al (2007) also distinguish between $\sigma_{12}(r)$ and $\sigma_{12}(\sigma)$.
They found that $\sigma_{12}(r)$ is larger than $\sigma_{12}(\sigma)$ (compare their Fig.4 with Fig.6). We have shown with our simulations that $\sigma_{12}(\sigma)$ can describe well $\sigma_{12}(r)$ provided $\pi_{max}$ is small enough.
Different groups have found this dependence of $\sigma_{12}$ on the scale (Zehavi et al 2002, Hawkins et al 2003, Jing and Borner 2004, Li et al 2006, Van den Bosch 2007, Li et al 2007 and citations there). 
It is difficult to provide a detailed comparison with these previous results. 
This is because very different assumptions are used: the infall model, the modeling 
of the real-space correlation, the values of $\pi_{max}$ or the use of Fourier 
versus configuration space analysis.
 However, these previous results seem to find lower values of $\sigma_{12}(\sigma)$ 
than our results, obtaining a maximum pairwise velocities of around 600 km/s, 
rather than then 800 km/s that we find. We believe that this difference is mainly caused by the methodology, which can make $\sigma_{12}(\sigma)$ lower that $\sigma_{12}(r)$ when large values of $\pi_{max}$ are used.
In our case we find with our modeling that smaller values of $\pi_{max}$ recover better the 
right values of $\sigma_{12}(r)$ when the galaxy bias $b$ is large, as in the case of LRG. 
These larger values at $\sigma_{12}(r)$ agree well with the dark matter prediction 
for $\Omega_m=0.25$ and $\sigma_8=0.85$, as found in paper I, so there is no
need to postulate that LRG velocities are in any way different.



\subsection{Consistency of $\sigma-\pi$ model: FOG}
\label{sec:consistency}

\begin{figure}
\centering{ \epsfysize=6cm\epsfbox{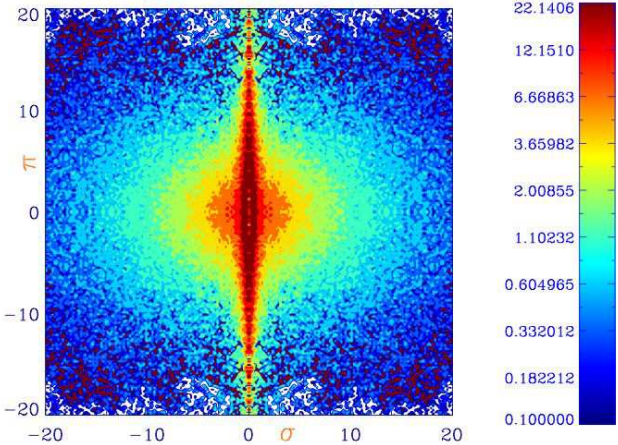}}
  \caption{Redshift space 2-point correlation function $\xisp$ calculated using squares in $\pi-\sigma$ 
  of side=0.2Mpc. Contours are: 0.1-50 with logarithm separation=0.6. 
  We can see clearly the fingers of God in the line-of-sight direction 
  \label{fig:pisigmasol}}
\end{figure}

Now we look directly at $\xisp$ at small scales once we have all the
parameters, to see if the model works when we separate $\pi$ and $\sigma$, for
all the angles, rather than just the monopole or quadrupole.
We have used a binning of 0.2Mpc/h for these plots, in order to
see clearly the fingers of God, which are concentrated at very small $\sigma$.
First, we can see the detailed plot of the measurements $\xisp$ in Fig.\ref{fig:pisigmasol}.

\begin{figure*}
\centering{ \epsfysize=10cm\epsfbox{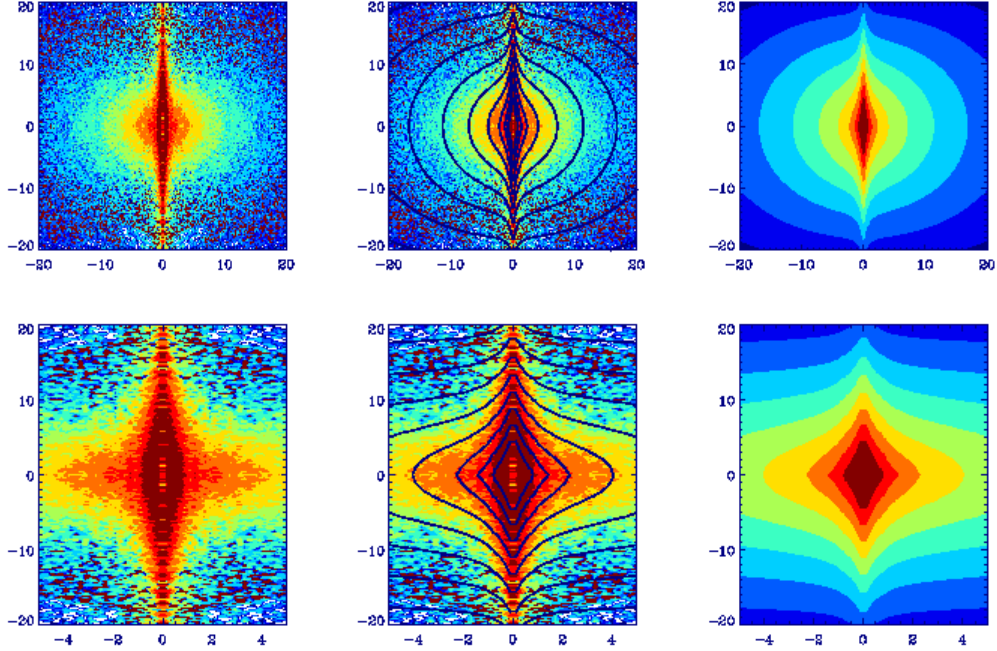}}
  \caption{Redshift space 2-point correlation function $\xisp$ (for the main slice in redshift z=0.15-0.47) modeled with a linear bias in the real-space correlation function, that we need to obtain $\xisp$ with Eq.(\ref{eq:hamiltonmethod}) and an effective $\sigma_{12}$=380km/s, as find for LRG at large scales. Top:data (as colors), data (as colors) + model (solid line), model (as colors). Bottom: the same zoomed in $\sigma$ \label{fig:pisigmabl}}
\end{figure*}

\begin{figure*}
\centering{ \epsfysize=10cm\epsfbox{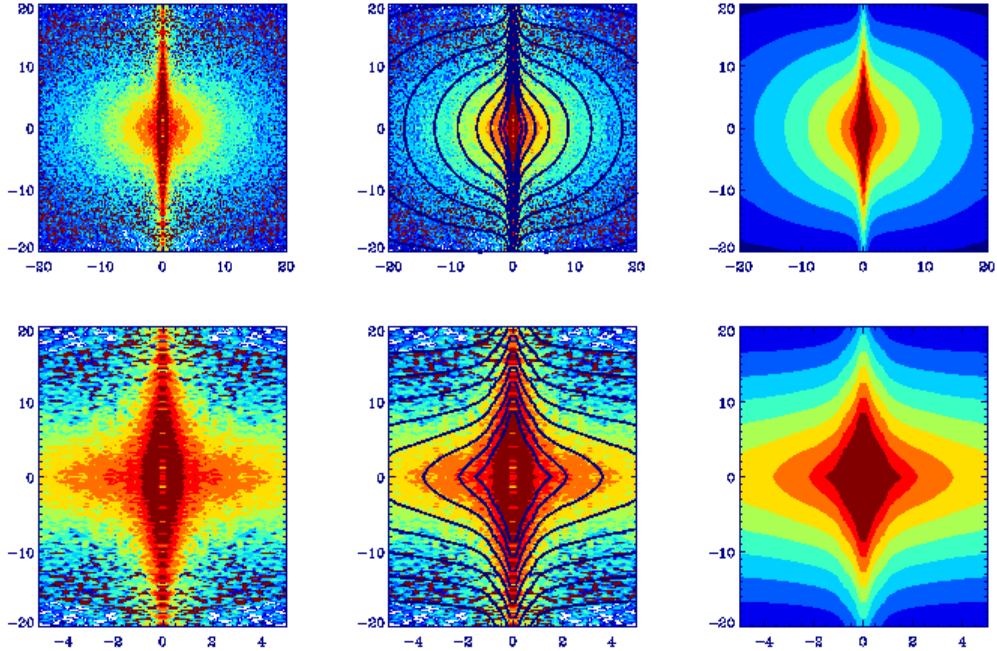}}
  \caption{Redshift space 2-point correlation function $\xisp$ (z=0.15-0.47) modeled with the real-space correlation function obtained from deprojection (this one has non-linear bias dependent on scale) and an effective $\sigma_{12}$=380km/s. We use Eq.(\ref{eq:hamiltonmethod}) to calculate the model. Top:data (as colors), data (as colors) + model (solid line), model (as colors). Bottom: the same zoomed in $\sigma$  \label{fig:pisigmar}}
\end{figure*}

\begin{figure*}
\centering{ \epsfysize=10cm\epsfbox{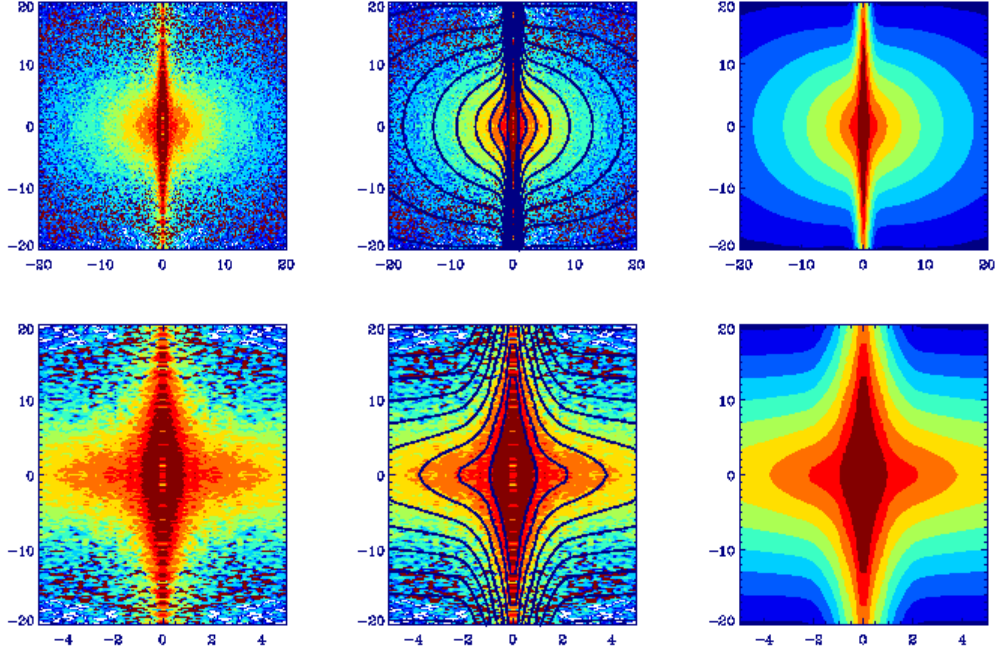}}
  \caption{Redshift space 2-point correlation function $\xisp$ (z=0.15-0.47) modeled with the real-space correlation function obtained from deprojection (using Eq.(\ref{eq:hamiltonmethod})) and $\sigma_{12}$ dependent on scale, as plotted in Fig.\ref{fig:realsigmavchange}. We include the variation of $\sigma_{12}$ with $\sigma$
and  assume that $\sigma_{12}$ is constant along the line-of-sight for a fixed $\sigma$ (perpendicular distance). Top:data (as colors), data (as colors) + model (solid line), model (as colors). Bottom: the same zoomed in $\sigma$  \label{fig:pisigmar2}}
\end{figure*}

\begin{figure*}
 \centering{ \epsfysize=6.cm\epsfbox{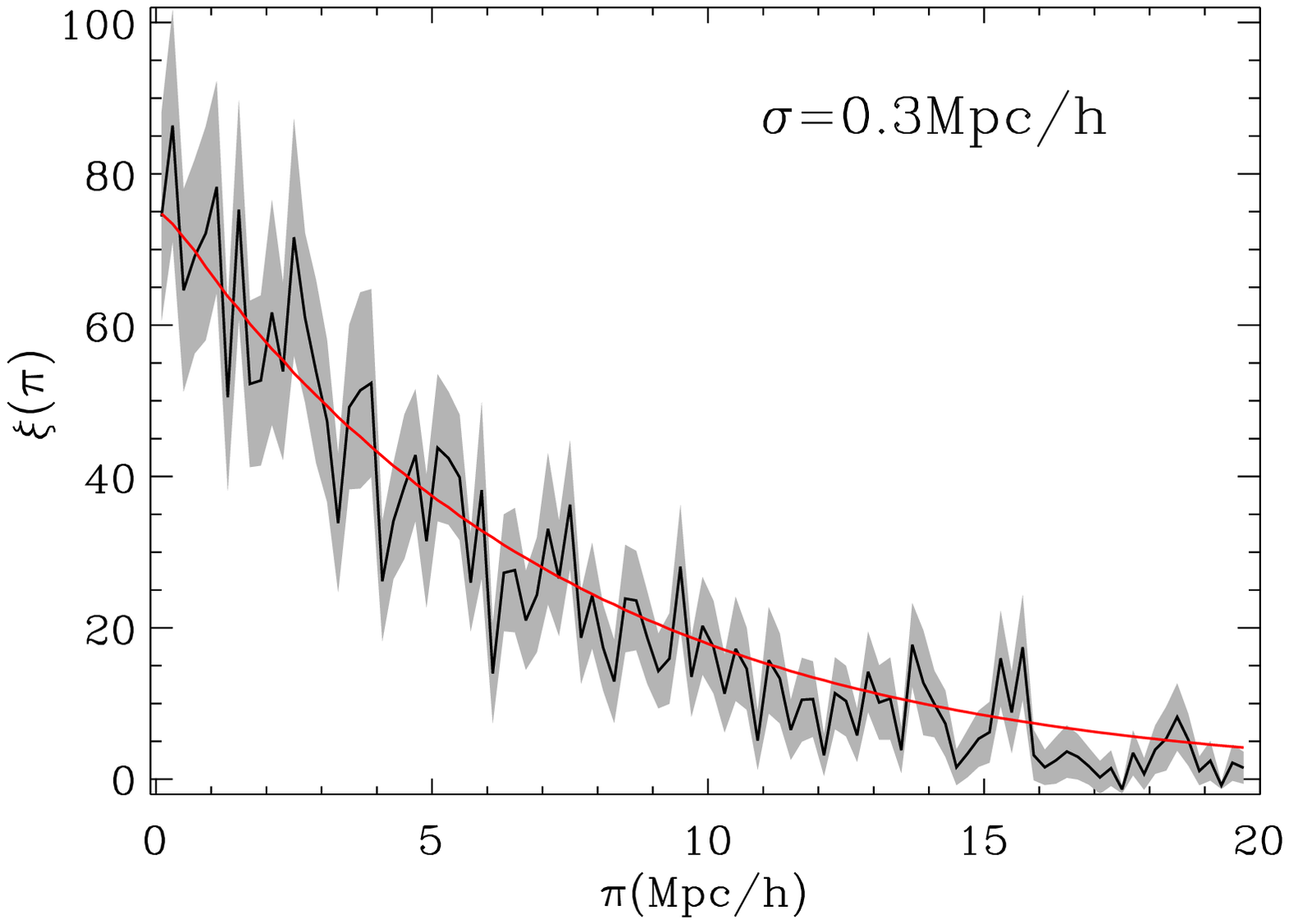}}
 \centering{ \epsfysize=6.cm\epsfbox{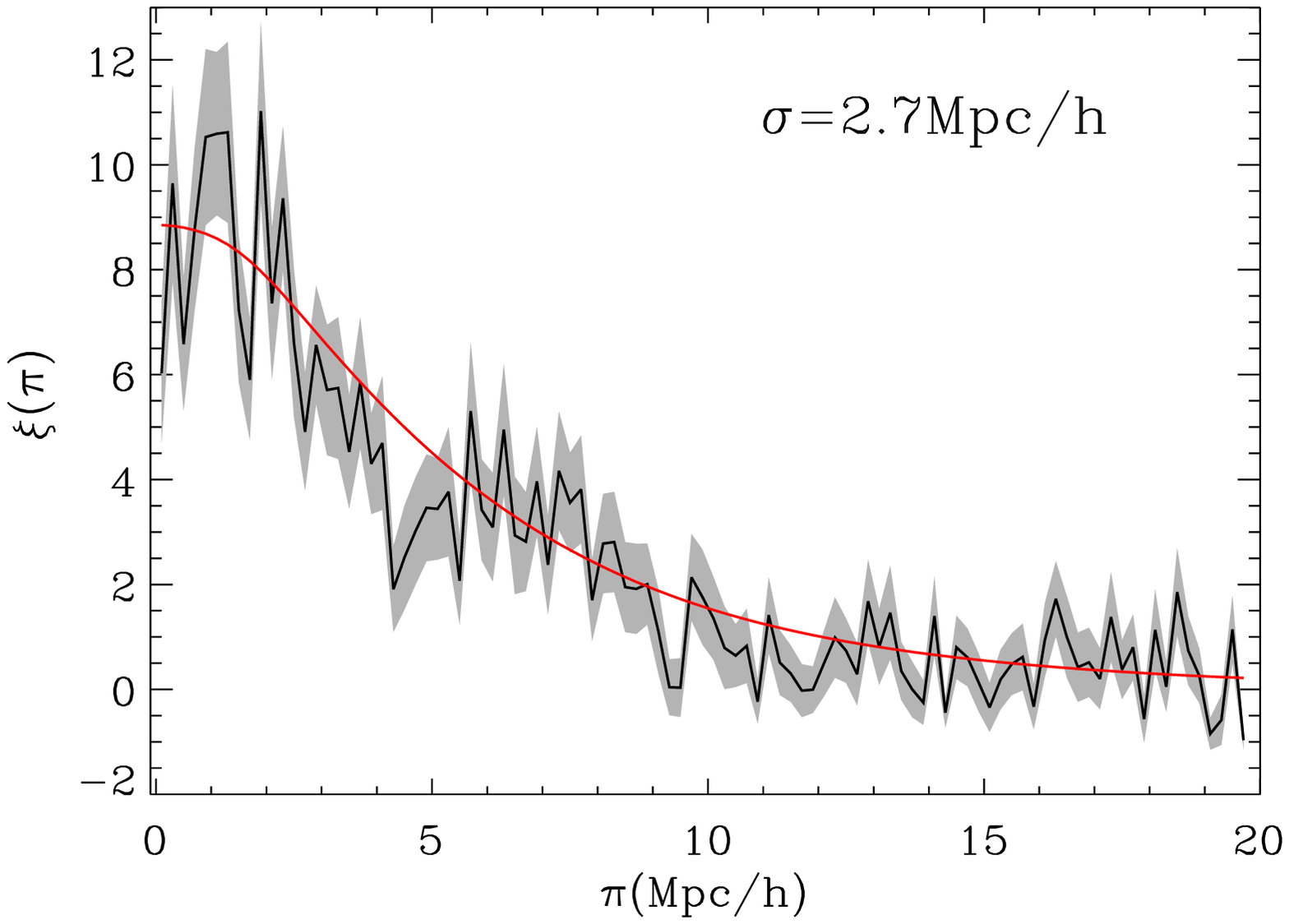}}
 \centering{ \epsfysize=6.cm\epsfbox{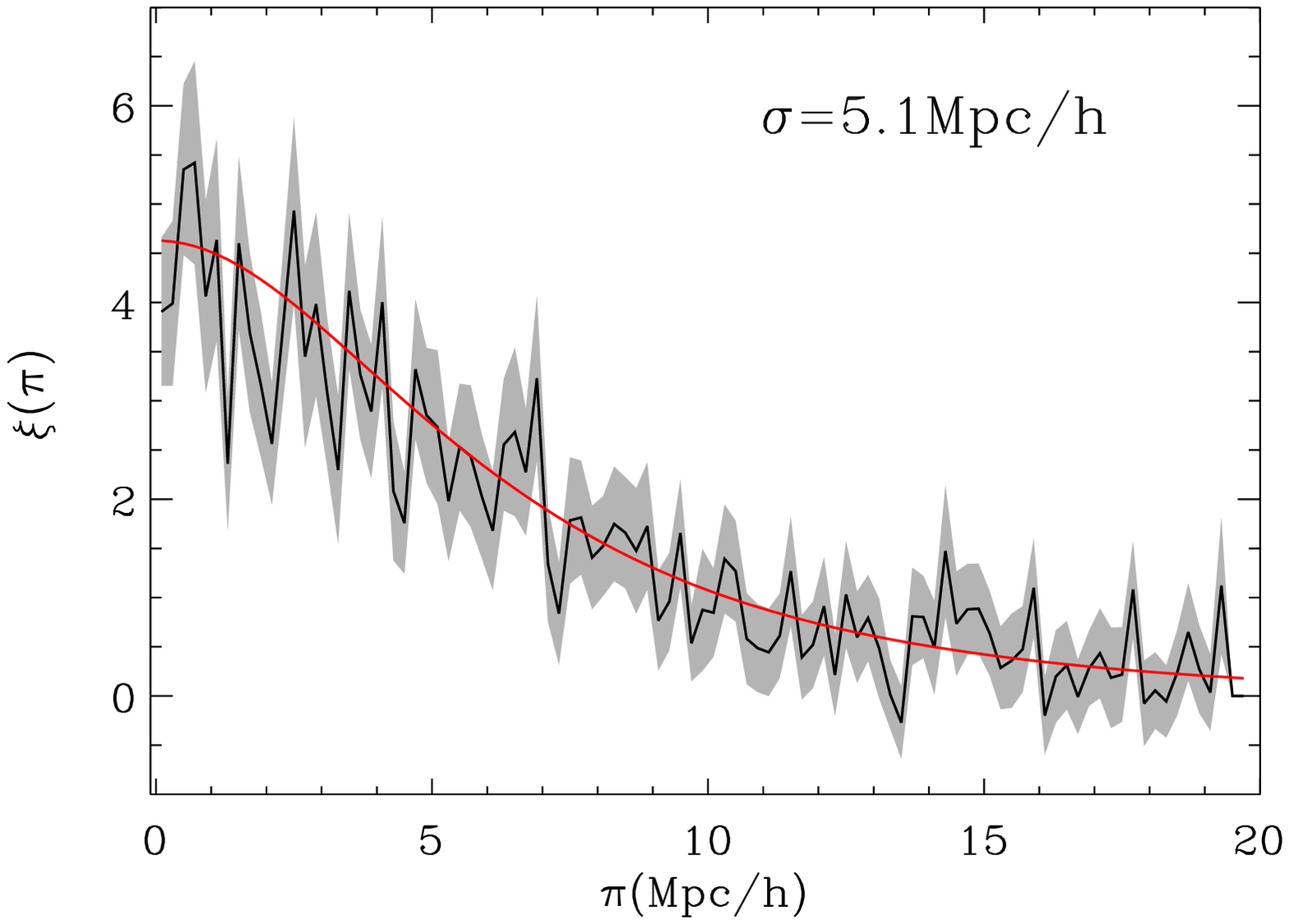}}
 \centering{ \epsfysize=6.cm\epsfbox{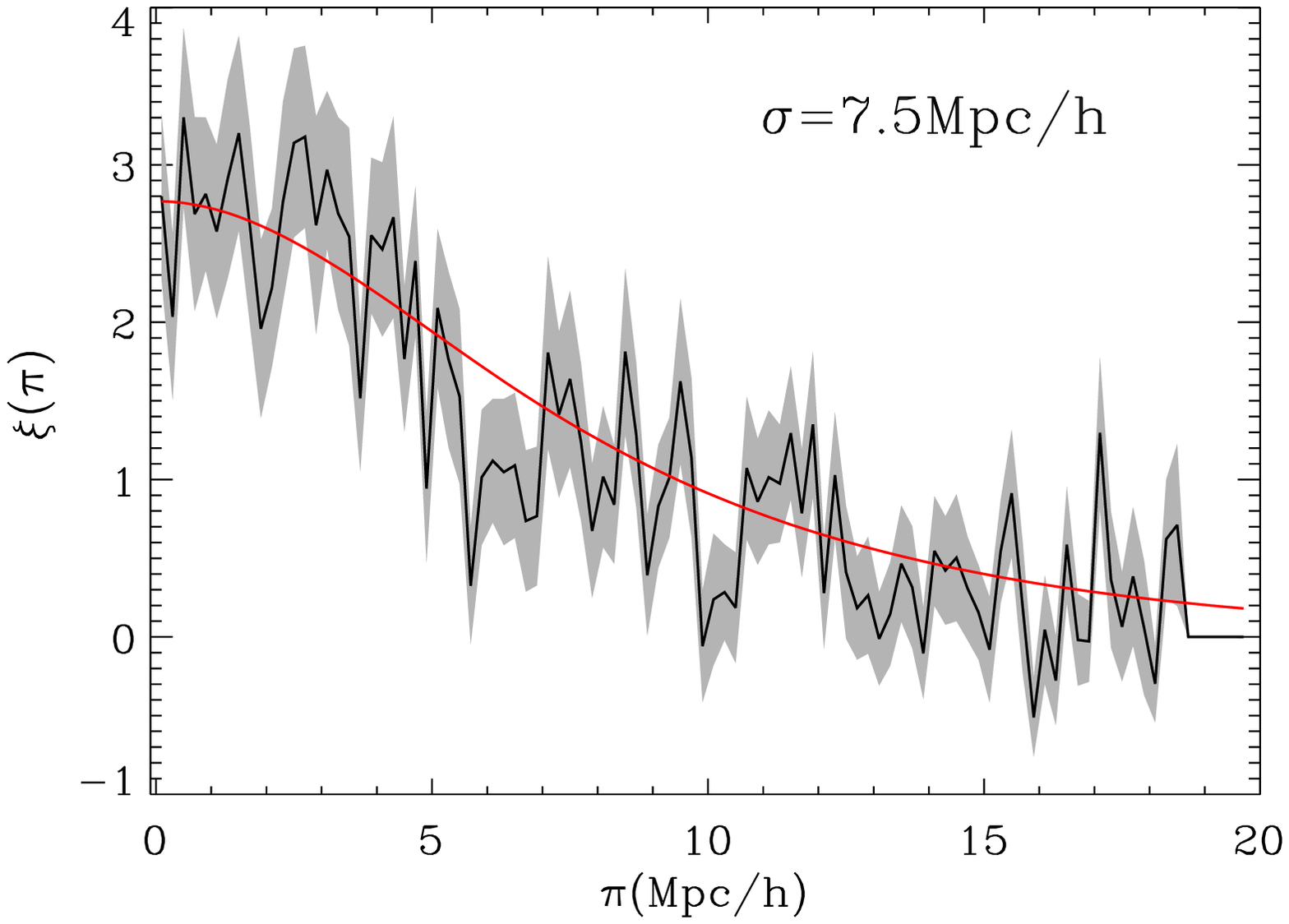}}

  \caption{We plot $\xisp$ as in Fig.\ref{fig:pisigmar2} along the line-of-sight $\pi$ for different fixed $\sigma$ in bins of 0.2 Mpc/h as labelled in the panels. Jagged solid line with errors (gray zone) is the measured LRG $\xisp$ while the smooth line (red) is the best model (with non-linear bias and $\sigma_{12}$ dependent on scale). Here we only show a few values of $\sigma$, but the model works well in all cases. \label{fig:pisigmar22}}
\end{figure*}

In the next three figures \ref{fig:pisigmabl}, \ref{fig:pisigmar} and
\ref{fig:pisigmar2}, we show the differences between the data and the model
in three cases. Top panels from left to right show: the data as colors; the data
and the model over-plotted as solid line; and the model as colors. Bottom panels
show: the same as in the top but we have zoomed the $\sigma$ direction to see
clearly the fingers of God.

In Fig.\ref{fig:pisigmabl}, we compare the data
with a model that assumes linear bias $b$ (found fitting large scales only) and a
constant pairwise velocity dispersion of $\sigma_{12}$ (obtained from the
quadrupole Q(s)). We see clearly that this does not work well. The FOG (along
the $\pi$ direction) are too small compare to data and the correlation
in the $\sigma$ direction has a different slope.
This is partially due to the fact that the
bias in the real data becomes non-linear on smaller scales.  
Part of the apparent lack of fingers of God  in the models are corrected just
adding the non-linear bias in the model.

In  Fig.\ref{fig:pisigmar}, we compare the model that we obtained using the real
space correlation function just found, which includes all the non-linear
effects, once we extrapolate $\xi(r)$ below 0.3Mpc/h, where fiber collisions prevent us from measuring clustering.
We can see that we still need to explain the strong elongation we see in the direction
line-of-sight, which we correct with the third model in Fig.\ref{fig:pisigmar2}.
Here we include the variation of $\sigma_{12}$ with scale, assuming that $\sigma_{12}$ is constant for large scales
and changes for small scales, as found in Fig.\ref{fig:realsigmavchange}. Here we use the exact modeling of $\sigma_{12}(r)$ as a function of $r$ as explained in section \ref{sec:model}. In order to see more clearly the validity of the model, we have plotted in Fig.\ref{fig:pisigmar22} the correlation function $\xisp$ along the line-of-sight $\pi$ for different fixed $\sigma$ as indicated in the figure. Similar
results are found for all other $\sigma$ values, which is studied in
bins of 0.2 Mpc/h. The model (smooth red solid line) seems to work well for small scales compared to the data (jagged lines with errors).

In the three cases, we can calculate the $\chi^2$ to see how well the models fit to the data, as 

\begin{equation}
\chi^2=\sum_{\pi,\sigma} (\xisp_{real}-\xisp_{model})^2/err(\pi,\sigma)^2
\end{equation}

using diagonal JK errors, which we have proved in the errors section that work for small $\sigma$.
We have seen that the covariance is small (less than 20\%) and so we
believe that this  $\chi^2$ estimation should be accurate (at least to
about 20\% accuracy). We see how the $\chi^2$ improves when we add a non-linear bias and variation of $\sigma_{12}$. 

We calculate the $\chi^2$ in different fitting zones, which we define by including in our analysis all the pixels that have an amplitude higher than 1.6, 2.4 or 4.4 in the third model, the one that is most similar to LRG (non-linear bias and variation of $\sigma_{12}$). The reduced $\chi^2$ (total $\chi^2$ divided by the number of individual points) varies from 2.2 to 3.2 for the first model
(linear bias) depending on the fitting zone; from 1.8  to 2.8 for the second model (non-linear bias);
and from 1.2 to 1.3 for the third model (non-linear bias and
variation in $\sigma_{12}$).
 The third model represents a major improvement respect to the other models although the fit
 is not perfect because the reduced $\chi^2$ is not equal to unity. 
However, the reduced $\chi^2$ remains constant when changing the fitting zone, 
so the model is consistent at different scales. Moreover, at small scales, a 
reduced $\chi^2$ of 1.2 is not rule out as can be seen in the first panel 
of Fig.\ref{fig:reducedchi2}. Apart from this effect,
we think that the difference to a perfect reduced $\chi^2$ can be attributed to the covariance which is small but non zero for LRG and has been neglected in this analysis. We should also take into account that in order to recover $\sigma_{12}$ we go through many steps, that can increase the error at the end. As an input for the calculation of the model of Eq.(\ref{eq:hamiltonmethod}) we need $\beta$ and the real-space correlation function $\xi(r)$, and both measures are estimations, they are not direct observables. 
We know that the model is perfect for dark matter, but it probably needs the inclusion of non-linearities to explain better the 2-halo pairwise velocity dispersions in highly biased galaxies as LRG (Scoccimarro 2004). We let to future work the inclusion of more realistic,
non-linear corrections, to the Kaiser model.
 As a conclusion, we see that this simple model can explain the strong FOG without need of having large pairwise velocities, and the model recovers almost all the features of $\xisp$, a complicated mix of different effects, as explained in previous sections.

\subsection{Different redshift slices}

\begin{figure}
\centering{ \epsfysize=6cm\epsfbox{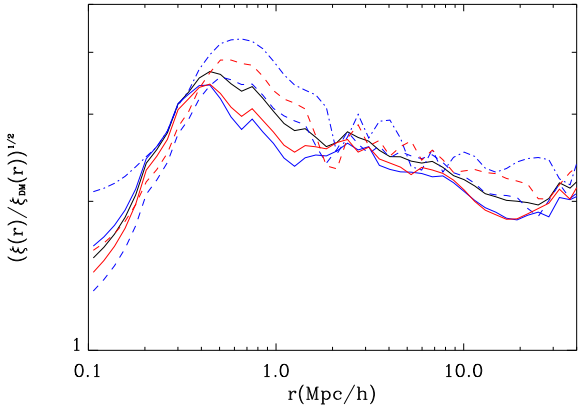}}
  \caption{Comparison between the bias $b(r)=\sqrt{\xir/\xir_{DM}}$ (as in Eq.(\ref{eq:bnolineal})) in real space for different slices in redshift. All: black, z=0.15-0.3 (solid blue), z=0.3-0.4 (dashed blue), z=0.4-0.47 (dashed-dotted blue); z=0.15-0.34 (solid red), z=0.34-0.47 (dashed red)  \label{fig:biasslices}}
\end{figure}

Finally, we look at the differences in the bias, defined in Eq.(\ref{eq:bnolineal}),
for different redshift slices. In Paper I we saw that
$b(z)D(z)$ was nearly constant as a function of redshift. 
This means that $b(z)$ grows smoothly with
redshift in agreement to what we see in Fig.\ref{fig:biasslices}. 
The slope in the non-linear bias
is nearly the same, so the small scale interaction between galaxies is nearly
the same as a function of redshift, as expected.\\

We have done the same analysis for different slices and the models
for $\xisp$ work very well for all the cases. Plots are very similar
to the ones in previous section.

\section{Discussion}    
\label{sec:discuss}

We have studied small scales in $\xisp$  measured from luminous red galaxies, 
which have both a strong non-linear bias and are affected by large random 
peculiar velocities in redshift space. 
We have compared our results for the monopole in redshift space $\xi(s)$, the real-space correlation function $\xi(r)$ and the perpendicular projected correlation function $\Xi(r)$ with Zehavi et al (2005) and Masjedi et al (2006) results. Our analysis agree at all the scales except for very small scales, below 1Mpc/h, where we find some minor differences with respect to previous analysis.

Here we also recover, for the first time,
the pairwise velocity dispersion $\sigma_{12}$ of LRG galaxies
as a function of separation by fitting $\xisp$.
Our method is shown to work in realistic simulations. We find that 
the scale variation of $\sigma_{12}$  in LRG galaxies is similar 
to that in  dark matter simulations. We show in
section \S\ref{sec:consistency} that a simple Kaiser model convolved with an 
exponential distribution of pairwise velocities, can explain 
well the complicate shape of $\xisp$ at small scales, once we 
add the scale dependent bias and the scale dependent $\sigma_{12}$. The $\chi^2$ per degree of freedom reduces by over a factor of two when we allow $b$ and $\sigma_{12}$ to change with scale. We notice that if we 
attribute all the distortion at small scales to the peculiar velocities, without taking into 
account the non-linear bias, velocities appear   to be artificially larger. 
On small scales the errors in $\xisp$ are relatively small compare
to the signal, see section \S\ref{sec:errorpisigma}.
Thus the agreement of the $\xisp$ data to our simple modeling shown in
Fig.\ref{fig:pisigmar2} is quite remarkable and significant. 
We believe that this is the first time this simple $\xisp$ model is shown to agree in detail with small scale LRG clustering.

\section*{Acknowledgments}

We would like to thank Pablo Fosalba, Francisco Castander, Marc Manera
and Martin Crocce for their help and support at different stages of
this project.
We acknowledge the use of simulations from the MICE consortium 
(www.ice.cat/mice) developed at the MareNostrum supercomputer
(www.bsc.es) and with support form PIC (www.pic.es),
 the Spanish Ministerio de Ciencia
y Tecnologia (MEC), project AYA2006-06341 with
EC-FEDER funding, Consolider-Ingenio CSD2007-00060
and research project 2005SGR00728
from Generalitat de Catalunya. AC acknowledge support
from the DURSI department of the Generalitat de
Catalunya and the European Social Fund.

\end{document}